%% file: dd87a11.tex
\documentclass[12pt,preprint]{aastex}
\usepackage{natbib}
\newcommand{\sna}{SN\,1987A}
\newcommand{\kms}{~{km\,s$^{-1}$}}
\newcommand{\amucc}{~{amu\,cm$^{-3}$}}
\newcommand{\utau}{~{s\,cm$^{-3}$}}
\newcommand{\tttt}[1]{{$\times 10^{#1}$}}
\newcommand{\pms}[2]{\,{$^{+{#1}}_{-{#2}}$}}
\newcommand{\hii}{H{\footnotesize \,II}}
\newcommand{\shii}{\mathrm{H\,II}}
\newcommand{\nvii}{N{\footnotesize \,VII}}
\newcommand{\oviii}{O{\footnotesize \,VIII}}
\newcommand{\nex}{Ne{\footnotesize \,X}}
\newcommand{\mgxii}{Mg{\footnotesize \,XII}}

\newcommand{\fexvii}{Fe{\footnotesize \,XVII}}

\newcommand{\sixiv}{Si{\footnotesize \,XIV}}
\newcommand{\chandra}{{\it Chandra}}
\newcommand{\xmm}{{\it XMM-Newton}}
\newcommand{\pha}{{\tt pha}}
\newcommand{\arf}{{\tt arf}}
\newcommand{\rmf}{{\tt rmf}}
\newcommand{\obs}{ObsId}
\newcommand{\tgcat}{{\it TGCat}}
\newcommand{\vb}{{v-b}}
\newcommand{\svb}{{\mathrm{v}\textrm{\footnotesize -}\mathrm{b}}}

\slugcomment{ ---  2012 April 19  ---  ApJ accepted. }

\shorttitle{Very-broad line component of \sna}
\shortauthors{Dewey et al.}

\begin{document}

\title{Evolution and Hydrodynamics of the Very-Broad X-ray Line
Emission in \sna}

\author{D. Dewey\altaffilmark{1}, V.V. Dwarkadas\altaffilmark{2}, 
F. Haberl\altaffilmark{3}, R. Sturm\altaffilmark{3},
C.R. Canizares\altaffilmark{1} }
\altaffiltext{1}{MIT Kavli Institute, Cambridge, MA 02139, USA;
dd@space.mit.edu}
\altaffiltext{2}{University of Chicago, Chicago, IL 60637, USA;
vikram@oddjob.uchicago.edu
}
\altaffiltext{3}{Max-Planck-Institut fuer extraterrestrische Physik,
Giessenbachstrasse, Garching, Germany D-85748
}

% Summarize concisely the content and conclusions.

% Not to exceed 250 words;
%   currently ``cat | wc'' gives: 245 (includes latex stuff)

\begin{abstract}
Observations of \sna\ by the \chandra\ High Energy Transmission Grating (HETG) in
1999 and the \xmm\ Reflection Grating Spectrometer (RGS) in 2003
show very broad (\vb) lines with a full-width at half-maximum (FWHM) of
order $10^4$\kms; at these times
the blast wave was primarily interacting with the \hii\ region around
the progenitor. 
Since then, the X-ray emission has been increasingly
dominated by narrower components as the blast wave encounters
dense equatorial ring (ER) material.
Even so, continuing \vb\ emission is seen in the grating
spectra suggesting that interaction with \hii\ region material is on-going. 
Based on the deep HETG 2007 and 2011 data sets, and confirmed by RGS
and other HETG observations, the \vb\ component has a
width of 9300\,$\pm2000$\kms~FWHM and contributes
of order 20\,\% of the current 0.5--2~keV flux.
Guided by this result, \sna's X-ray spectra
are modeled as the weighted sum of the
non-equilibrium-ionization (NEI) emission from two simple 1D
hydrodynamic simulations, this
``2$\times$1D'' model reproduces the observed radii, light curves, and spectra with
a minimum of free parameters.
The interaction with the \hii\ region
($\rho_{\rm init}\approx 130$\amucc, $\pm$\,15~degrees
opening angle) produces
the very-broad emission lines and most of the 3--10~keV flux.
Our ER hydrodynamics, admittedly a crude approximation to the
multi-D reality, gives ER densities of
$\sim$\,$10^4$\amucc, requires dense clumps ($\times$5.5
density enhancement in $\sim$\,30\,\% of the volume),
and it predicts that the 0.5--2~keV flux will drop at a rate of
$\sim$\,17\,\% per year once no new dense ER material is being shocked.
\end{abstract}


% max of 6; in alpha order.
\keywords{ Hydrodynamics --- ISM: supernova remnants --- radiation mechanisms: thermal
--- supernovae: individual: SN 1987A --- Techniques: Spectroscopic ---
X-rays: general}



%%%%%%%%%%%%%%%%%%%%%%%%%%%%%%%%%%%%%%%%%%%%%%%%%%%%%%%%%%%%%%%%%%%%%%%%%%%%%%
\section{INTRODUCTION}
\label{sec:intro}

Since its explosion as the first optical SN observed in the year 1987,
\sna\ has been the subject of intense study at every possible
wavelength. Its location in the Large Magellanic Cloud (LMC), 
at a relatively close distance of 51.4\,$\pm$1.2~kpc \citep{Panagia99},
has enabled us to
study details regarding the aftermath of a SN explosion
not previously possible; we have learned and will
continue to learn more about the physics of SNe and the formation
of supernova remnants (SNRs).

The X-ray emission from \sna\ has likewise been studied with every
available satellite. However, \sna\ is by far the intrinsically dimmest X-ray SN to be
observed, at least in its first decade, and is only now approaching
the luminosity of other comparable-age X-ray SNe \citep{Dwarkadas12}.
It is clear that the only reason it was detected in early
observations at high X-ray energies (6--28~keV) with the Ginga satellite \citep{Inoue91}
was because of its proximity. The emission appeared to decrease
over the next couple of years, and it appeared as though \sna\ would
behave like every other X-ray SN then known, with an X-ray emission
decreasing with time. 

Just over 3 years after its explosion however, the X-ray
\citep[0.5--2~keV ROSAT observations]{Hasinger96} and radio
\citep[1.4--8.6~GHz with the Australian Telescope Compact Array]{Gaensler97}
emission from \sna\ began instead to increase with time. The
increase in radio emission was initially interpreted by
\citet{Chevalier92} as arising from the interaction of the SN forward
shock (FS)
with the wind termination shock of the blue supergiant (BSG) wind
of the progenitor. However this would not result in a continued
increase but would instead be limited in time. The observations
indicated otherwise, and the X-ray emission continued to
increase.
The subsequent increases have suggested that the interaction is
with a much higher density and more extended region than expected from the wind
termination shock. This was interpreted by \citet{Chevalier95} as
ionized red-supergiant (RSG) wind emitted during a pre-BSG phase,
and subsequently swept-up by the fast BSG wind.
The ejecta interaction with the resulting \hii\ region
produces the usual two-shock structure with the FS moving
into the \hii\ region and a reverse shock (RS) moving back
into the ejecta; a contact-discontinuity (CD) defines the ejecta-\hii\
boundary.  Such a model was used to explain the early X-ray emission
\citep{BBMcC97let} and predicted line widths of 5000\kms\ or more.

\begin{figure*}[t]
\begin{center}
\includegraphics[angle=270,scale=0.55]{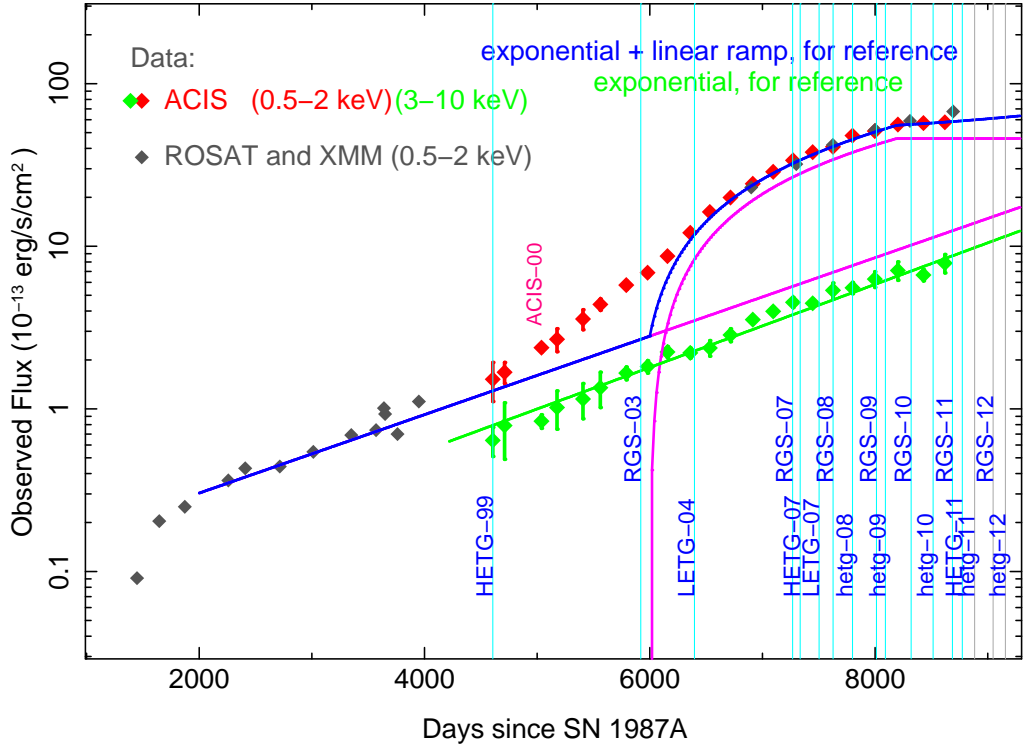}  \\
\end{center}
\caption{\sna\ X-ray light curves with grating observation epochs indicated.
The 0.5--2~keV data (black, red) show an initial exponential behavior whose
slope increases after day $\sim$\,4500 and then decreases
from day $\sim$\,6300, possibly leveling off at the
recent epoch $\sim$\,8800 days.
In constrast, the 3--10~keV flux (green) has more closely followed a
consistent exponential behavior to the present.
For reference, the sum of an exponential (beginning at day 2000)
and a linear ramp increase (from day
6000 to day 8200), is included (magenta, blue) to highlight the general phases of the
0.5--2~keV light curve.  The ROSAT fluxes are taken from \citet{Haberl06}
and the ACIS values are from \citet{Park11}.
\label{fig:lcoverview}}
\end{figure*}

The X-ray emission since then has continued to rise, as shown in
Figure 1 for the 0.5--2~keV and 3--10~keV bands.
A review of the X-ray emission over the first 20 years is
given in \citet{McCray07}. The hard X-ray and radio emission have
continued to rise together, but after about 6000 days the soft X-ray
light curve has steepened still further, indicating a source of
emission that leads predominantly to soft X-rays. This has been
interpreted by \citet{McCray07} as due to the FS interacting with
dense fingers of emission arising from the inner edge of the
equatorial ring (ER) surrounding \sna.  
From days 6500 to 8000 the soft X-ray flux continued to grow
at a roughly linear rate\footnote{This was first pointed out by D.N. Burrows in
discussions during the preparation of \citet{Park11}.},
 that is $dF/dt\approx$ constant.
Most recently, \citet{Park11} have reported a flattening of the X-ray light curve
since day $\sim$\,8000,
indicated in Figure~\ref{fig:lcoverview}, and have suggested
the possibility that the FS has now propagated beyond the majority
of the dense inner ring material.  This could lead to further
changes in the light curve and, hence, future measurements are eagerly
anticipated.

Even considering its low luminosity, the proximity of \sna\ has
allowed for high resolution grating observations to be taken.
These observations, including recent ones from 2011, are
summarized and analyzed in \S\,\ref{sec:gratobs}.
There is much to learn from this set of high-resolution X-ray spectra;
however, here we focus on the existence and evolution of a ``very-broad''
(\vb) emission line component in the data, \S\,\ref{sec:vb}.
This \vb\ component is persistent over the last decade 
and it is likely the continuation of the broad lines expected by \citet{BBMcC97let} and 
measured in the \chandra\ HETG data of \citet{Michael02}.
In \S\,\ref{sec:hydro} we review and present our hydrodynamic modeling
of \sna\ which consists of two contributions:
the \hii\ material above and below the equatorial plane
and the dense equatorial ring material.
We compare the results of the model with observed quantities 
in \S\,\ref{sec:results},
showing that the \vb\ component naturally arises from the shocked 
\hii\ material and is the main source of the 3--10~keV flux.
Also, our very simplified hydrodynamics of the ring
interaction is used to show how the 0.5--2~keV light curve
would behave if the FS has indeed gone beyond the densest ring material.
In the final section we summarize our conclusions.

% light curve with grating epochs indicated
% created by:  /Users/Dan/_Science/SN1987A/isis_120201/hylc.sl

\clearpage
%%%%%%%%%%%%%%%%%%%%%%%%%%%%%%%%%%%%%%%%%%%%%%%%%%%%%%%%%%%%%%%%%%%%%%%%%%%%%%
\section{GRATING OBSERVATIONS AND DATA REDUCTION}
\label{sec:gratobs}

%% In a manner similar to \objectname authors can provide links to data set
%% hosted at participating data centers via the \dataset{} command.  The
%% second curly bracket argument is printed in the text while the first
%% parentheses argument serves as the valid data set identifier.  Large
%% lists of data set are best provided in a table (see Table 3 for an example).
%% Valid data set identifiers should be obtained from the data center that
%% is currently hosting the data.
%%
%% Note that AASTeX interprets everything between the curly braces in the 
%% macro as regular text, so any special characters, e.g. "#" or "_," must be 
%% preceded by a backslash. Otherwise, you will get a LaTeX error when you 
%% compile your manuscript.  Special characters do not 
%% need to be escaped in the optional, square-bracket argument.

The grating observations of \sna\ are listed in Table~\ref{tab:gratobs},
including those from instruments on \chandra~(High- and Low- Energy Transmission
Gratings: HETG, LETG) and \xmm~(Reflection Grating Spectrometer, RGS).
The observation epochs are marked on the \sna\ light curve of
Figure~\ref{fig:lcoverview}, showing that only the
HETG-99 observation \citep{Michael02} was taken before day 5500
after which the flux from the shock--protrusion collisions became significant.
Along with HETG-99, the deep non-grating \chandra\ ACIS-00 observation
at 5036 days \citep[\obs\,1967]{Park02} gives the best pre-collision X-ray spectra (albeit at
only CCD resolution) and is included in the table as well.
With the collision getting underway, the RGS-03 \citep{Haberl06} and LETG-04
\citep{Zhekov05,Zhekov06} observations cover a transition period.
Since 2007, with most of the X-ray flux due to the shock-ring
interaction, there have been regular grating observations
made with both \chandra\ \citep{Dewey08,Zhekov09} and \xmm\ \citep{Sturm10}.

For our data analysis, we use standard spectral extractions from the
observations as described in \S\,\ref{sec:data} \&
\S\,\ref{sec:rgsdata}.
The various analyses,
described in \S\,\ref{sec:3shock} \& \S\,\ref{sec:vb},
are then carried out in ISIS \citep{Houck00}
using a common set of scripts that include some simple tailoring
of the energy ranges and spatial-spectral parameters (see
Appendix~\ref{sec:gsmooth})
based on the type of data, e.g., HETG vs.\ RGS.

\subsection{\chandra\ Data}
\label{sec:data}

Public \chandra\ grating data and their spectral extractions are
available and were obtained from the \tgcat\ archive\,\footnote{\,
{\tt http://tgcat.mit.edu/}}
\citep{TGCat11}.
All of these had been re-processed in 2010 or later and therefore the
downloaded \pha, \arf\ and \rmf\ files
could be used without any further re-processing.
Multiple \obs's at the same epoch were combined in
ISIS by summing the \pha\ counts, creating
an exposure-weighted \arf, and summing the exposure times.
To reduce background from the less sensitive front-illuminated CCDs,
the \pha\ and \arf\ for wavelengths longward of 17.974\,\AA\ were set
to zero in the plus [minus] order of the MEG [LEG] data.
The plus and minus orders are then combined, even though
they do have small but significant
differences in their line-widths because of
the resolved nature of \sna\ \citep{Zhekov05,Dewey08}.
The composite line-shapes are accounted for in the model by a smoothing
function that has parameters based on the {\it observed}
line widths, \S\,\ref{sec:3shock} and Appendix~\ref{sec:gsmooth}.
In this way we reasonably approximate the combined-orders' line shapes.
Thus, further analysis is
carried out on a single spectrum in the case of LETG data
and on two spectra (MEG and HEG) for HETG observations.

Some HETG observations are indicated with a lower-case ``hetg-YY''
designation in the table.
These spectra are from the combined ``monitoring observations'' (PI -
David Burrows) in the year 20YY.  Starting in 2008 the HETG was
inserted for \sna\ observations in order to reduce the
effects of pileup \citep{Park11}; as a by-product, additional HETG high-resolution data
are taken.
However, to obtain the shortest exposure times only a subframe of the ACIS
is read out; because of this the outer portions of the
HETG dispersed ``X'' pattern \citep{Canizares00,Canizares05}
fall outside of the subframe, removing response at longer wavelengths.
Instead of centering \sna\ in the subframe, it is offset
in cross-dispersion to get more of the 
sensitive MEG minus order in the subframe and this extends the
response to just include the \oviii\ line at $\sim$\,0.65~keV.

The most recent HETG observations of \sna, HETG-11 in
Table~\ref{tab:gratobs}, were carried out during 2011 March 1--13
as part of the GTO program (PI - Claude Canizares).
The proprietary data were retrieved from the
\chandra\ archive and scripts from \tgcat\,\footnote{\,See
  Help$\Rightarrow$Software on the \tgcat\ web page:
{\tt http://tgcat.mit.edu/}} were used to carry out the spectral extraction
using standard CIAO tools; specifically
CALDB 4.4.5 and CIAO 4.3 were used to process the data. 

The non-grating ACIS-00 observation, \obs\,1967, was downloaded from the
archive and processed through level 1 event filtering with CIAO tools.
The {\tt psextract} routine was then used to generate \pha, \arf,
and \rmf\ files for a $\sim$\,6 arc second diameter region around \sna;
a background extraction was made but not used because it had very
few counts.

\subsection{\xmm\ RGS Data}
\label{sec:rgsdata}

The European observatory, \xmm, has
been used to study \sna\ at high spectral resolution via
its reflection-grating spectrometer, RGS \citep{denHerder01}.  The data and
processing steps for the four epochs of
RGS data taken through January 2009 have been presented in \citet{Sturm10}.
Two subsequent observations, RGS-10 and RGS-11,
have been carried out and were similarly processed;
\xmm\ SAS 10.0.0 was used in (re-)processing all RGS data.
When read into ISIS
the RGS-1 and RGS-2 counts (and backgrounds) are kept separate and jointly fit
in further analysis.

% - - -
\input{tab_gratobs.tex}

% - - -

\clearpage
%%%%%%%%%%%%%%%%%%%%%%%%%%%%%%%%%%%%%%%%%%%%%%%%%%%%%%%%%%%%%%%%%%%%%%%%%%%%%%
\subsection{Fluxes and Three-Shock Model Fitting}
\label{sec:3shock}

The \chandra\ grating observations can robustly determine the observed
flux (and statistical error) in energy bands directly
from the data without using a model.
Instead of the the usual coarse ``flux in the 0.5--2~keV band,''
fluxes in Table~\ref{tab:gratobs} are given for 
three bands covering this range.  The band boundaries are chosen to
avoid strong lines and are dominated by lines of N\,\&\,O,
Fe\,\&\,Ne, and Mg\,\&\,Si, respectively.  Because of their reduced
wavelength coverage the ``hetg-YY'' data sets do not allow an accurate flux
measurement in the lowest-energy band.  For the RGS data sets
the fluxes given in the table are from 3-shock model fits (described
below) because the RGS spectral gaps and background counts complicate
a direct-from-the-data approach.

Comparing HETG-11 fluxes to the previous HETG-07 ones shows
that the main deviation from the trend of continued flux increase
has happened primarily in the lowest energy range:
while the 0.79--1.1 and 1.1--2.1 keV fluxes have increased by factors
of 1.78 and 2.17, respectively, the 0.47--0.79 keV flux has
increased by a smaller factor of only 1.18.\footnote{\,The apparent
low-energy decrease may have a contribution due to calibration issues,
in particular with the ACIS contamination modeling.
An HETG observation of the SNR E0102 was taken
on 2011 February 11, just a
month before the \sna\ data; its initial analysis shows
calibration changes in the oxygen lines, but only of order 20\,\%.}
Likewise, similar spectral changes are seen
in the recent RGS data of 2010 December (``RGS-11''),
continuing trends seen earlier; for example, 
whereas the \nex\ flux increased by 36\,\% from RGS-08 to RGS-09
the lower-energy \nvii\ flux increased by only 10\,\% 
in the same period \citep{Sturm10}.

A nominal non-equilibrium ionization (NEI) model was fit independently to each
epoch's data with the goal of capturing all of the relevant
lines in that epoch's spectrum.  In order to allow for the wide range of shock
temperatures seen in \sna\ \citep[Figure~2]{Zhekov09},
the model consists of the sum of three plane-parallel shock components 
with common (``tied'') abundances.  Physically, a single plane-parallel shock
model represents the emission from a uniform slab of material into which a
shock wave has propagated for some finite time, $t_s$ \citep{Borkowski01}.
The shocked portion of the slab has uniform electron temperature, $kT_e$,
and density, $n_e$, but has a linear variation in the time since shock passage, 
going from $\sim$\,0 immediately postshock to a maximum of $t_s$ for the
earliest-shocked material.  Thus, the plane-parallel shock emission can be
parameterized by a normalization ($\propto n_e n_H V$), the abundances, 
the postshock electron temperature, and the maximum ionization
age\,\footnote{\,Specifically, the maximum ionization age is specified by the {\tt
Tau\_u} parameter of the {\tt vpshock} model in XSPEC; there is also
a {\tt Tau\_l} parameter which in keeping with our physical description is set to 0.}, 
$\tau=n_e t_s$.

The sum of the three shock components is then smoothed with a Gaussian function to account for
line-broadening and the finite size of \sna.
Finally, there is an overall photo-electric absorption term to account
for Galactic and LMC absorption.
The model is implemented in ISIS, making use of the XSPEC spectral
models library; the model specification (Expression~\ref{eq:3shock}) and
a description of the smoothing parameters
are given in Appendix~A.
Many of the model parameters are fixed as in previous
work \citep{Zhekov09}: the column density\,\footnote{\,The
column density could have a measureable local component
that would change as the \sna\ system evolves: taking a
density of 1\tttt{4}~H\,cm$^{-3}$
with a 0.01~pc path length sets an $N_{H\,{\rm local}}$ scale of $\sim$\,0.3\tttt{21}.
Clumping, the ionization state, and the geometry of \sna\ have to be taken into account as well.
In terms of data analysis, the $N_H$ is very degenerate with the low-energy
line abundances (N, O) and
with the continuum teperatures and norms.
For our purposes we fixed $N_H$
at the two-shock-model value determined in \citet{Zhekov09},
roughly half of which is due to the Galactic contribution of $\sim$\,0.6\tttt{21}\,cm$^{-2}$.
}
($N_H=$\,1.3\tttt{21}\,cm$^{-2}$),
the redshifts (286.5\kms), and the abundances of elements
without strong, visible lines (H=1, He=2.57, C=0.09, N=0.56, Ar=0.54,
Ca=0.34, Ni=0.62); as previously the solar photospheric abundances
of \citet[Table 2]{AG89} are used as the reference set.
There remain the free parameters
consisting of 3 normalizations, 3 temperatures, 3 ionization ages,
and the abundances of O, Ne, Mg, Si, S, and Fe.

The parameter set was reduced further by fixing the middle temperature at a value 
near the (logarithmic) center of the emission measure distribution of \sna,
$kT_{\rm \,mid}=$\,1.15~keV \citep[Figure~2]{Zhekov09}.  
Since the $\sim$\,1~keV emission had grown from
2004 to 2007, the expectation, borne out in
Tables~\ref{tab:3shock}\,\&\,\ref{tab:3shVB}, is that this component
would have a growing contribution to the model, especially at later epochs.
To test the (in)sensitivity of the fit values to the exact value
of $kT_{\rm \,mid}$, the HETG-11 data were fit with additional
$kT_{\rm \,mid}$ values of 1.0\,\&\,1.3~keV.  All of the HETG-11 fit parameters
showed small variations over the test set of mid-temperature values,
\{1.0, 1.15, 1.3\} keV. Some examples are: 
$kT_{\rm \,lo}$ takes values \{0.50, 0.54, 0.56\},
the $kT_{\rm hi}$ norm varies as \{2.4, 2.0, 1.5\},
the silicon abundance varies as \{0.41, 0.39, 0.38\}, and
the $\chi^2$ changed slightly as \{1069, 1088, 1111\}
(for 560 degrees of freedom.)
These small, gradual variations demonstrate that the exact choice 
for the middle temperature is somewhat arbitrary and nearly degenerate
with other parameters.

Two more parameters are removed by constraining the ionization ages
to cover a factor of 2, e.g., as seen in \citet{Zhekov09},
specifically setting:
$\tau_{\rm \,lo}=\sqrt{2}\,\tau_{\rm \,mid}$ and
$\tau_{\rm \,hi}=\tau_{\rm \,mid}\,/\sqrt{2}$.
Hence, in addition to the abundances there are 6 free parameters:
$kT_{\rm \,lo}$, $\tau_{\rm \,mid}$, $kT_{\rm \,hi}$, and
the 3 normalizations, $N_{\rm lo}, N_{1.15}, N_{\rm hi}$.  This is the same
number of parameters as for a general 2-shock model (2 norms, 2 $kT$'s, 2 $\tau$'s) and gives
slightly better fits.

% Made with isis_120201/plot_3sh_comps.sl
\begin{figure*}[t]
\begin{center}
\includegraphics[angle=270,scale=0.65]{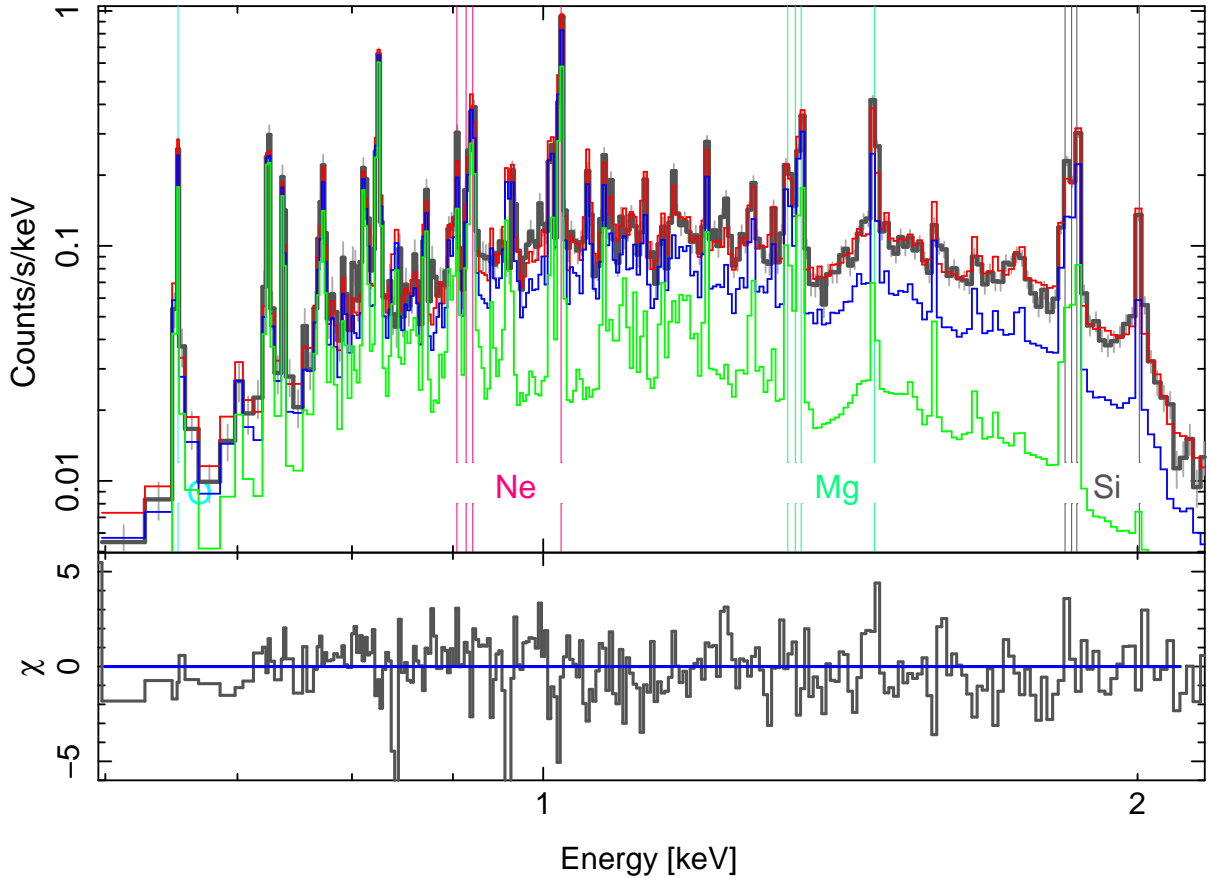}  \\
\end{center}
\caption{HETG-11 data and model.  The data
(black) are globally fit by a 3-{\tt vpshock} model (red).
Removing the high-temperature component gives the blue curve and
just the low-temperature component is shown in green. 
\label{fig:3shock}}
\end{figure*}

The 3-shock model was fit to the HETG and LETG data in the 0.6 to 5 keV
range, giving good overall fits, Figure~\ref{fig:3shock}.
For fitting, the spectra were binned to a minumum of 20--30 counts per bin and a
minimum bin width of $\sim$\,0.05~\AA.
The HEG data are ignored below 0.78~keV ($>$15.9\,\AA) and ignored
completely for the HETG-99 data.  
Because of their reduced wavelength coverage,
several changes were made for the ``hetg-YY'' data sets:
they were ``noticed'' in the range 0.69--5~keV for the MEG (1.25--5~keV for
the HEG), the O abundance was fixed, and the $kT_{\rm \,lo}$
value was fixed (these latter two were set to values based on the HETG-07
and HETG-11 fits.)
The RGS data sets do a good job of covering the N line $\sim$\,0.5~keV
but have reduced sensitivity above the \sixiv\ line, 
so these data were fit in the range 0.47 to 2.20~keV with the N
abundance free, the S abundance frozen at a nominal value (0.36), and
the $kT_{\rm \,hi}$ value fixed.
The ACIS-00 spectrum was fit over the 0.4 to 8.1~keV range with N free.

The values for the 3-shock fit parameters and their
1-$\sigma$ confidence ranges are given in Table~\ref{tab:3shock}.
As expected from their low fluxes, the two pre-collision data sets (HETG-99,
ACIS-00) show comparatively low normalizations.
Note that the RGS values and the HETG/LETG values differ significantly at similar
epochs; this is probably due to {\it near-degeneracies in the model}
which amplify the differences between the spectrometers in terms of their
energy-range coverage, statistical weighting (counts vs.\ energy), and
line-spread functions.  However they do show the same general trends,
e.g., toward larger $\tau_{\rm \,mid}$ values at more recent epochs.
In the next section we use these 3-shock fits as good starting points
to look for and measure the presence of the \vb\ component.

% Put the table at the end of the section
\clearpage
\input{tab_3shock.tex}

\clearpage
%%%%%%%%%%%%%%%%%%%%%%%%%%%%%%%%%%%%%%%%%%%%%%%%%%%%%%%%%%%%%%%%%%%%%%%%%%%%%%
\subsection{The Very-Broad Component and Its Evolution}
\label{sec:vb}

% Figure(s) created with isis_120201/plot_vb_resid.sl :
\begin{figure*}[t]
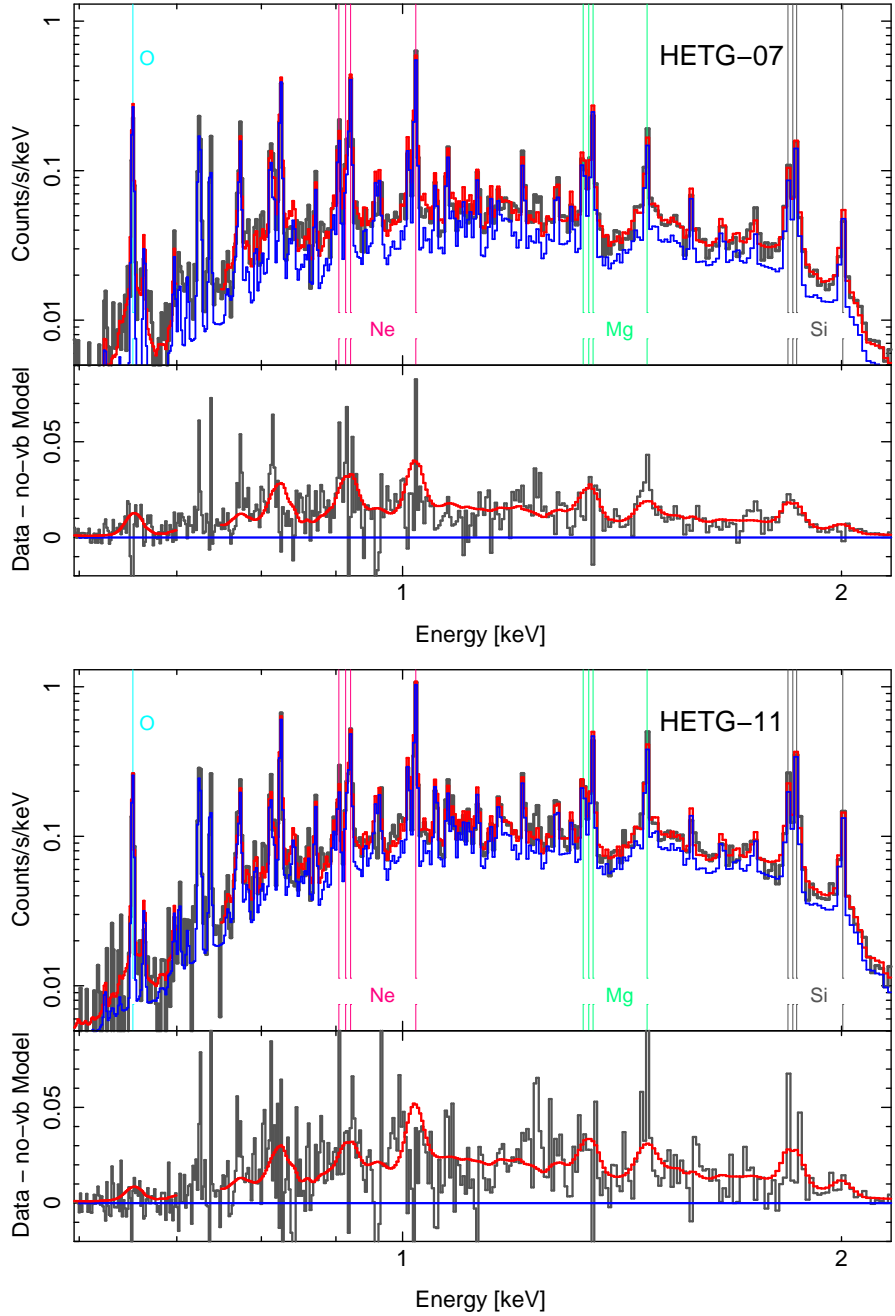

\begin{center}
\includegraphics[angle=270,scale=0.48]{HETG-07_vbresid.ps}  \\
\vspace{0.1in}
\includegraphics[angle=270,scale=0.48]{HETG-11_vbresid.ps}  \\
\end{center}
\caption{Deep HETG spectra and their very-broad (\vb) models.
The HETG-07 (top pair) and HETG-11 (bottom pair) data are shown (black)
with the best-fit \vb\ 3-shock model (red).  The non-broad
model component is shown as well (blue).  The lower half plots
show the difference (black) between the data and the non-broad component
with the \vb\ component over plotted (red).
\label{fig:vbresid}}
\end{figure*}

To look for and quantify a \vb\ line component in the \sna\ spectra
we start with the data and 3-shock model fits (above) and extend the
model to include a \vb\ component that is a broadened
version of the 3-shock model itself; the model definition is explicitly
given in Appendix~A, Expression~\ref{eq:vb}.  The \vb\ extension
adds two additional parameters: the fraction of the total flux that is in the \vb\
component ($f_\svb$) and the width of the \vb\ component expressed
as an equivalent FWHM in velocity ($v_\svb$). 

The choice of broadening the whole spectrum is the simplest given the
low signal-to-noise ratio of the \vb\ component in any given line, provided
there is some expectation that the bright lines of the \vb\ component
are similar to those of the narrow emission.  At face value this may
seem to be {\it unlikely}, with the \vb\ lines produced by a plasma
with very different temperatures and ionization ages from the plasma
producing the narrow lines.
However, our hydrodynamic modeling shows that the approximation is a
surprisingly good one as seen in Figure~\ref{fig:vbhydro} and
discussed in \S\,\ref{sec:vbspectrum}.
However, based on this comparison
we have chosen to ignore the \fexvii\ 0.70--0.75~keV range
when doing \vb\ fitting (below) as it is the only
bright line-complex that is not predicted to have a \vb\ component.

The \vb--3-shock model is fit to the data using $\sim$\,60\%
finer binning and a somewhat smaller energy range that
focusses on the brightest, high-resolution lines in the particular spectra,
generally those of O, Ne, Mg, Si, and Fe-L.
For HETG, LETG and RGS data these ranges are, respectively: 0.62--2.1~keV, 0.6--1.6~keV
(no Si),
and 0.47--1.1~keV (including N but not Mg and Si).
For ``hetg-YY'' data sets because of their limited low-energy range,
the O line is not included and the O abundance is frozen
at the average of the HETG-07 and HETG-11 O \vb\ values.
Because of the reduced energy range of the \vb\ fitting, the $kT_{\rm \,lo}$, 
$\tau_{\rm \,mid}$ and $kT_{\rm \,hi}$ parameters were
fixed at their 3-shock values; this leaves the 3 normalizations and the
abundances to be fit along with the \vb\ fraction and width.

As examples, the \vb\ fits to the HETG-07 and HTEG-11 data are shown in
Figure~\ref{fig:vbresid}; note that the y-axes of the lower, difference
plots are the same, showing that the \vb\ flux is {\it increasing}
in absolute terms.
The statistics of this \vb\ fitting are given in the left-half of
Table~\ref{tab:vbstats}; in particular the values of $F_2$ are large
enough to indicate that the addition of the two \vb\ parameters has
improved the model fit significantly.

\clearpage

% Combined contours of the very broad parameters:
% created by:  /Users/Dan/_Science/SN1987A/isis_120201/vbconfplot.sl
\begin{figure*}[t]
\begin{center}
\includegraphics[angle=270,scale=0.50]{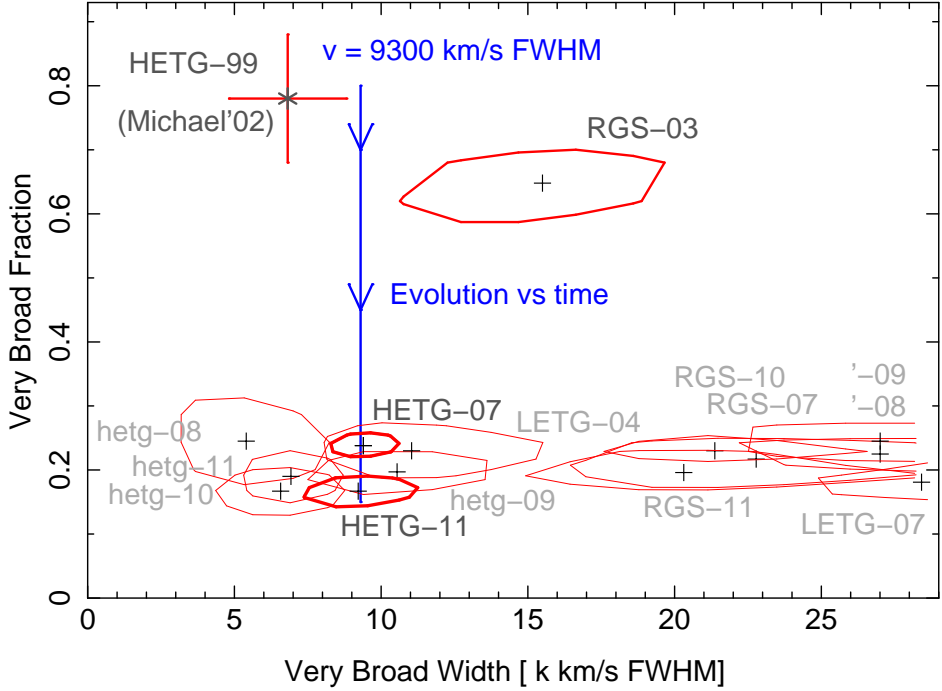}  \\
\end{center}
\caption{Confidence contours for the very-broad component
parameters.  Solid curves give the
68\,\% ($\Delta\chi^2=2.3$) contours for fits to 
each grating data set; the best-fit values are indicated with ``$+$'''s.
The two deep HETG observations, HETG-07 \& HETG-11,
show a similar, well-determined  \vb\ width of $\sim$\,9300\kms~FWHM.
See the text for a discussion of the spread seen among the contours.
\label{fig:vbcontours}}
\end{figure*}

Instead of tabulating the best-fit \vb\ parameters and their ranges, it is more
useful to generate two-dimensional confidence contours
in ``\vb\ fraction vs.\ \vb\ width'' space.
These contours and their best-fit values are shown in
Figure~\ref{fig:vbcontours}; for HETG-99 the data point with errors
is based on the 2-Gaussian ``stacked fitting'' result given in \citet{Michael02}.
The earliest data sets, HETG-99 and
RGS-03, show large values of \vb\ fraction ($>$\,0.6) with widths of
order 10\tttt{3}\kms.  In contrast the later
observations show \vb\ fractions within the range 0.15 to 0.30 and 
cover a large spread in widths, with the HETG values below
15\tttt{3}\kms\ and RGS-determined widths above this value.

\begin{figure*}
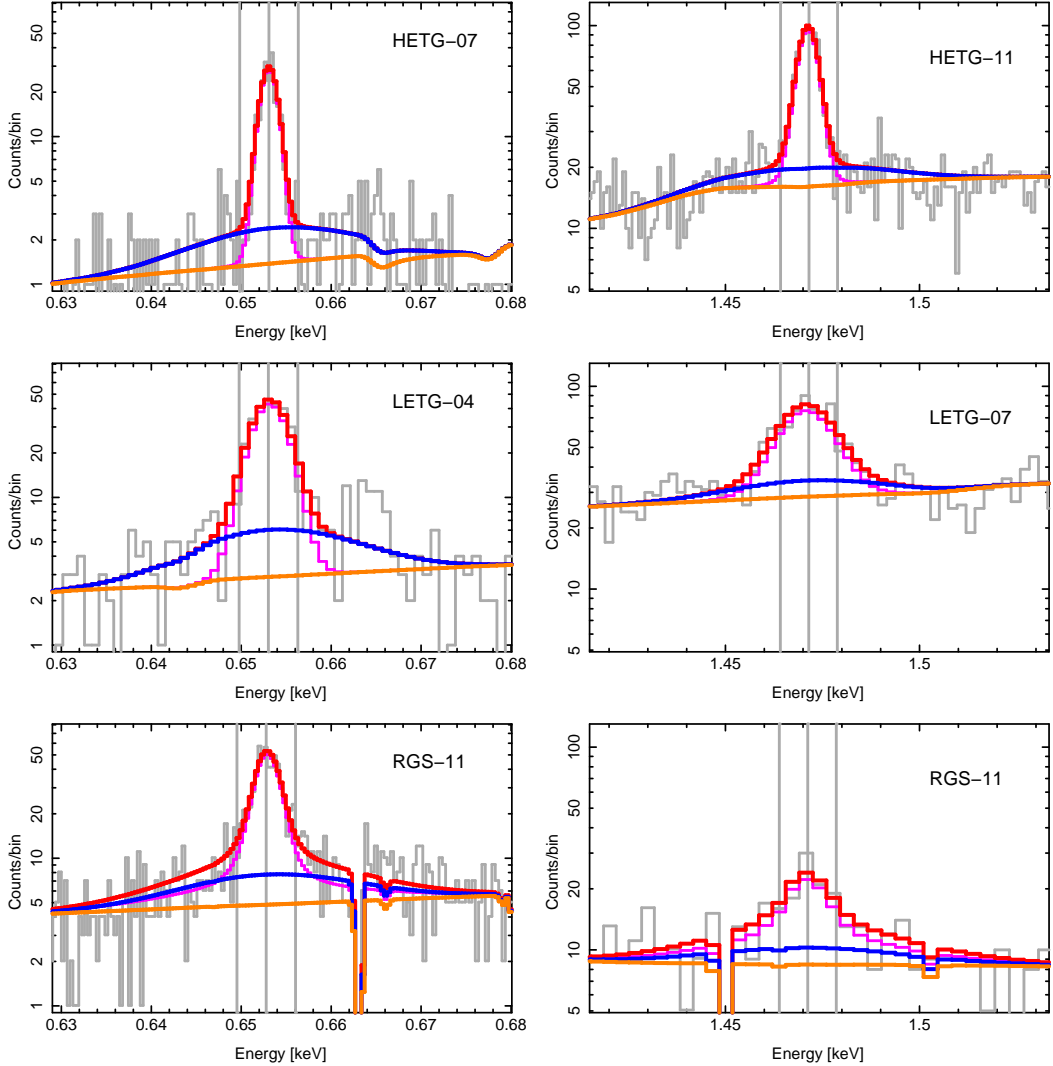

\begin{center}
\includegraphics[angle=0,scale=0.6]{plots_comp_oviii.ps}
\includegraphics[angle=0,scale=0.6]{plots_comp_mgxii.ps}  \\
\end{center}
\caption{Comparing the grating spectrometer line responses when observing \sna.
The HETG(MEG), LETG, and RGS views of the same combination of a
narrow line plus a very-broad
component are shown for the case of $f_\svb =25$\% and
$v_\svb =9300$\kms\ FWHM.
Because the resolution varies with energy, responses are shown
for \oviii\,(left colum) and for \mgxii\,(right column).
On each plot the orange curve shows a continuum level, the blue
curve is the continuum plus the \vb\ component, the magenta curve is the
continuum plus the narrow component, and the red curve is the sum of 
all three.  Vertical gray lines indicate the line center and
$\pm$\,1500\kms\ from it.
For reference, example data are shown in the background by the light gray 
histograms; the \oviii\ region includes the He-like L-$\beta$ line at
$\sim$\,0.665~keV.  The HETG provides the best separation of the \vb\ and narrow
components especially at the higher energies.
\label{fig:vbLSFs}}
\end{figure*}

This separation suggests that the measurement of the \vb\ component depends on the
particular spectrometer used.   Figure~\ref{fig:vbLSFs} shows how
the spectrometers ``see'' a monochromatic line consisting of a narrow
and a \vb-component; these plots include \sna's spatial-spectral
blur descibed in Appendix~\ref{sec:gsmooth}.
The HETG, even at the higher \mgxii\ energy, clearly separates
the narrow and \vb\ components, although it's also clear that the
\vb\ presence is subtle.
For the LETG the narrow component is significantly broadened by
the spatial extent of \sna\ and at high energies the \vb\ component is
almost completely covered by the narrow component.
The RGS line-spread function is more peaked than the LETG
at \oviii, however it includes extended wings so that even the
narrow component produces counts far from the line center.
Hence the RGS \vb\ measurement will be particularly sensitive
to the accuracy of its LSF calibration.

With these considerations in mind, we see that the two deep
HETG observations, HETG-07 and HETG-11,
should give the most accurate measure of the \vb\ component.
Comparing their small confidence contours,
there is a clear change in the \vb\ fraction from 2007 to 2011.
In terms of the \vb\ width, these two HETG contours show very
similar values of $\approx$ 9300\kms~FWHM, with an estimated 
90\,\% confidence range of $\pm$\,2000\kms.
This value is also in rough agreement with the (lower
statistics) HETG-99 and RGS-03 widths and so it is reasonable to
postulate a constant \vb\ width with a fraction that changes
over time as shown by the
blue path in Figure~\ref{fig:vbcontours}.

% Figure(s) created with isis_120201/comp_oviii.sl :

% The vb fitting Table
\input{tab_vb9300.tex}

Given the above, we fix the width of the \vb\
component at $v_\svb=9300$\kms~FWHM and fit for just the \vb\ fraction.
The results of this ``constrained'' fitting are given in the right-half of
Table~\ref{tab:vbstats}, including the
\vb\ fractions and their 68\,\% confidence ranges.
Note that for most data sets the $\chi^2$ change when doing this
constrained fitting, $\Delta\chi^2_1$, is a substantial portion of
the $\chi^2$ change when the \vb\ width is also being fit, $\Delta\chi^2_2$.
This suggests that the data do not strongly constrain the width to
differ from our assigned value;
the elongated contours of Figure~\ref{fig:vbcontours} suggest this as well.
Conversely, the \vb\ fractions given by the constrained
fitting have only a small dependence on the assigned width.
As examples, when the data sets LETG-07, RGS-11, HETG-11, and hetg-11 were
fit with the width set to 6500\,\&\,13000\,\kms~FWHM, the best-fit fractions
found are:  0.132\,\&\,0.160, 0.216\,\&\,0.210, 0.163\,\&\,0.178, and
0.200\,\&\,0.188, respectively.
These variations are actually within the statistical
1-$\sigma$ ranges given for the $f_\svb$ values in
Table~\ref{tab:vbstats}, and so the choice of the \vb\ width
will not significantly change the \vb\ fractions and light curve.

Having determined the \vb\ fractions for the data sets we can now
re-do the 3-shock fitting of \S\,\ref{sec:3shock} with
the appropriate \vb\ component fixed and included in the model.
This gives the results in Table~\ref{tab:3shVB}.
Note that the abundances have increased compared to the non--\vb\ values
in Table~\ref{tab:3shock}.
This is expected since a larger flux in a given line is now needed in
order to match the height of the narrow-line component in the high-resolution spectrum.
The change is especially pronounced
for HETG-99  where the \vb\ fraction is 0.78 and the abundances
are now in better agreement with those of ACIS-00 (although there is a large error range.)
Also as expected, the ACIS-00 values hardly change when
the \vb\ component is included because the lower-resolution CCD spectra
are primarily determined by the total line flux with little
sensitivity to any underlying velocity structure.

% The vb-3shock results
\input{tab_3shVB.tex}

\clearpage
%%%%%%%%%%%%%%%%%%%%%%%%%%%%%%%%%%%%%%%%%%%%%%%%%%%%%%%%%%%%%%%%%%%%%%%%%%%%%%
\section{HYDRODYNAMIC MODELING OF  \sna}
\label{sec:hydro}

Besides being an extremely well-studied object, \sna\ has been
frequently modelled in almost every wavelength regime. The
earliest calculations, which delineated the circumstellar material
(CSM) into a low-density
inner and high-density outer region, were made by \citet{Itoh87}. The
X-ray emission was modelled by \citet{Masai91} and
\citet{Luo91a} using 1-dimensional (1D) models. Two-dimensional (2D) models were
explored by \citet{Luo91b} for the case of an hour-glass equatorial
ring (ER) structure.
\citet{Suzuki93} carried out 2D axisymmetric smoothed particle
hydrodynamics (SPH) simulations with the ER modeled as a torus;
their Figures 4--8 demonstrate the hydrodynamic complexity by showing
the different components' spatial distributions.
\citet{Masai94} went back to the 1D models but took into
account the light-travel times from different parts of the remnant.
\citet{Luo94} carried out 2D calculations approximating the ER
as a rigid boundary; they also introduced the ejecta density distribution that
has been subsequently used by most authors, and forms the basis for
the ejecta density profile used in this paper.

The increasing X-ray and radio emission led \citet{Chevalier95} to
propose the existence of an ionized \hii\ region interior to the ER. The
FS had collided with this region around day 1200, and was
expanding in this moderately-high density medium,
$\rho\approx 100$\amucc, giving rise to increasing
X-ray and radio emission. The slowing down of the shock wave observed
at radio wavelengths was consistent with this conjecture. The X-ray
flux from the \hii\ region was modelled by \citet{BBMcC97let}, whereas a
detailed prediction of the expected X-ray flux and spectrum when the
shock collided with the dense ER was given in
\citet{BBMcC97}.

For the next decade or so, investigations of the X-ray emission were
mainly made via simpler, non-hydrodynamical models that attempted to
tie in the observed X-ray emission to the CSM properties
\citep{Park04,Park05, Park06, Haberl06, Zhekov10}.
\citet{Dwarkadas07RevMex} revised the older
simulations made (but not presented) in \citet{Chevalier95} using
more recent data, and managed to create spherically symmetric
hydrodynamical models that successfully reproduced the FS
radius and velocity as measured at radio frequencies. Using
these simulations he was also able to make a crude estimate of the
X-ray light curve. This model was updated in \citet{Dwarkadas07AIP}, and forms
the basis of the hydrodynamical simulations that are presented in the
following.

\subsection{Our Hydrodynamic Model of \sna}

% Schematics of Zhekov(2010) and this work at t=0 and t=7300 days.
% From keynote file:  SN1987A/diag_120314/z10_compare.key
\begin{figure*}[t]
\begin{center}
\includegraphics[angle=0,scale=0.33]{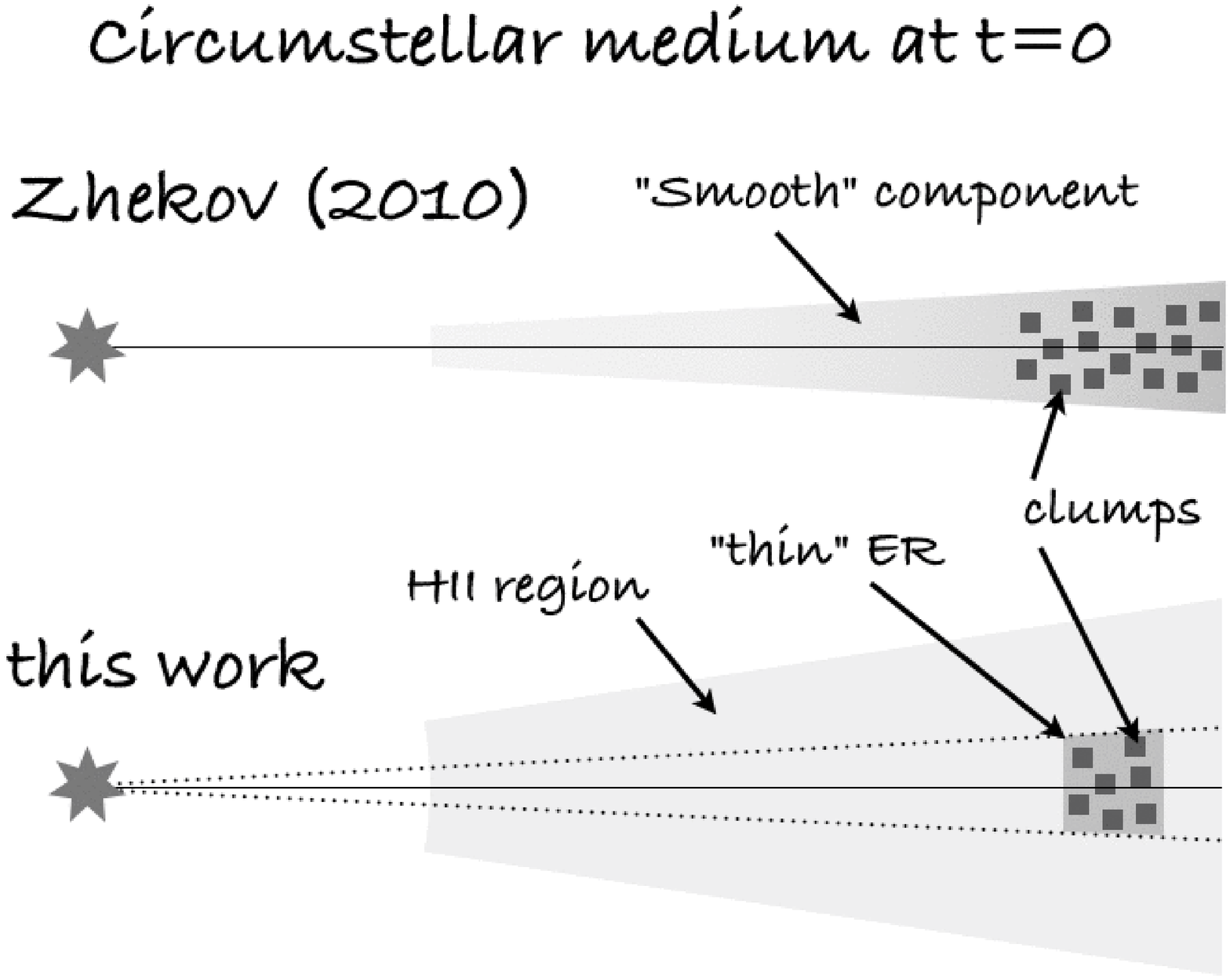}
\hspace{-0.1in}
\includegraphics[angle=0,scale=0.33]{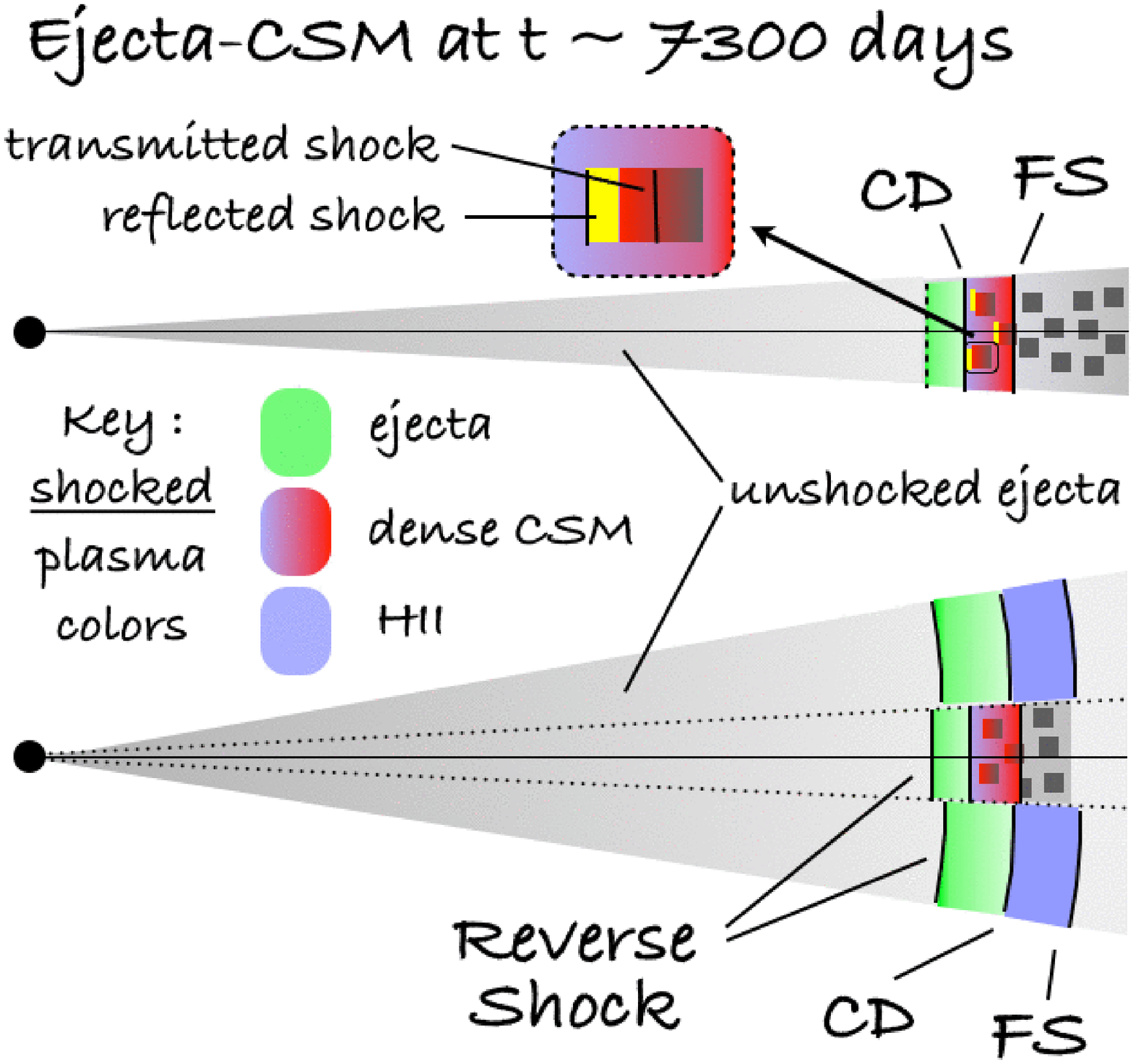} \\
\end{center}
\caption{Schematic diagrams of the CSM as modeled in
\citet{Zhekov10}, top row, and in this work, bottom row.  The left diagrams show the
CSM configuration assumed at the time of the SN explosion.
The coresponding diagrams at the right show the configuration
$\sim$\,7300~days after explosion when the forward shock, driven by
the expanding ejecta, has progressed into the dense, clumpy ER.
The Zhekov model is based on a single 1D hydrodynamics whereas our
model consists of two 1D models, one that includes the dense ER and
one that accounts for the \hii\ material above and below the
equatorial plane.  It is this latter component that gives rise to
the very-broad X-ray emission.
\label{fig:rsscompare}}
\end{figure*}

The structure of the CSM around \sna\ is inherently 3-dimensional, as is
visible in the amazing images from the {\it Hubble Space Telescope},
most recently in \citet{Larsson11}.
Modelling the ejecta-CSM interaction in 3-dimensions
is a highly challenging and computationally demanding problem,
does not encourage iterative parameter exploration, and requires
multi-dimensional non-equilibrium ionization (NEI) X-ray
emission calculations.
However, given the substantial symmetries in the overall geometry,
it is possible to partition the domain into one or more
regions whose emission can then be approximated by a
spherically symmetric 1D simulation.
These simulations can then be summed together, weighted
by the fraction of the full $4\pi$ solid angle that they each subtend.

As examples of this approach we show schematics in
Figure~\ref{fig:rsscompare} for the modeling of \citet{Zhekov10} and
our model.  The former consists of a single 1D solution with a smooth,
increasing-density CSM component.  To match the observed spectra, especially at
late times, dense ``clumps'' are included in the CSM with the clump
X-ray emission based on transmitted and reflected shock equations \citep{Zhekov09}.
Our model in the lower portion of Figure~\ref{fig:rsscompare}
is motivated by the on-going existence of the \vb\ component which
requires a combination of two 1D solutions:
\begin{itemize}
\item[i)] A ``with ER'' simulation that consists of an \hii\ region
followed at larger radii by a density jump to high values
at the ER location; the density profile is shown in
Figure~\ref{fig:initprofile}. This component subtends a small solid
angle, $\Omega_{\rm ER}$, and the high-density ER includes
additional ``clumping''
(\S~\,\ref{sec:lcs}\,\&\,\ref{sec:erabunds}).
This component is similar to the 1D model of \citet{Zhekov10}.
\item[ii)] A ``no ER'' simulation in which the \hii\ region continues
to large radii and there is no ER density jump.
This simulation produces the \vb\ emission and
corresponds to material above and below the
equatorial plane; it  subtends a moderate solid angle,
$\Omega_\shii$.
\end{itemize}
\noindent This multi-1D technique does have its
limitations, especially for the ``with ER'' case,
and we will keep these in mind.

The simulations here were carried out using the VH-1 code, a
1, 2, and 3-dimensional finite-difference hydrodynamic code based on
the Piecewise Parabolic Method \citep{Colella84}. The code solves the
hydrodynamical equations in a Lagrangian framework, followed by a
remap to the original Eulerian grid. Radiative cooling is included in
the form of the X-ray cooling function scaled by a factor of 3.5 to
account for IR dust emission, $L_\mathrm{IR}/L_\mathrm{X}\approx 2.5$. 

For each of a set of time steps, the code outputs a snapshot file giving
the hydrodynamic quantities density, velocity, and pressure at each
radial zone; two examples are shown in Figure~\ref{fig:modprofile}.
The set of snapshot files is then processed by custom routines coded
in S-Lang\,\footnote{\,S-Lang web page: {\tt http://www.jedsoft.org/slang/}}
and run within the ISIS software environment.  The
radial grid is rebinned into mass cuts, so that the time
evolution of each parcel of gas in a given mass shell can be
accurately followed; in this way the NEI state is tracked and
the X-ray emission can be calculated \citep{Dwarkadas10}.

Of course we did not simply pick parameters, run the simulation, and
have agreement with observations.  An iterative process of trial\,\&\,error
was carried out to converge on model parameters that reproduced the
various measured properties of \sna, especially in the X-ray.
Rather than take the reader through these steps (and mis-steps), we
summarize in the following sections the final model properties and
then compare the model results to data in \S\,\ref{sec:results}.

\subsection{Ejecta Parameters}
The ejecta are modelled using the \citet{Chevalier82selfsim}
prescription where the ejecta are in homologous expansion with a density 
that decreases as a power-law with radius, i.e.,
\begin{equation}\label{eq:ejecta}
\rho_{\rm ej} = C~ t^{-3}\,(r/t)^{-n}~.
\end{equation}
\noindent This is written to show that one can also consider the density profile as
a function of $v=r/t$.  Since a power-law extending all the way back to the origin
is unphysical, below a
certain velocity, $v_t$, the density is assumed to be a constant,
see the ``Ejecta'' portion of the density in Figure~\ref{fig:initprofile}.
Thus the profile can be specified by the three parameters: $C$, $n$, and
$v_t$ . The ejecta's total kinetic energy and mass
are functions of these parameters \citep[equ.\,2.1]{Chevalier89}
and so we can equivalently choose other 3-parameter sets, e.g.:
$n$, $E_{\rm ej}$, and $M_{\rm ej}$ .

In our 1D hydrodynamic
simulations, the RS moves into the ejecta and at most shocks
only the outer $\sim$\,0.5
solar masses of the ($4\pi$) ejecta profile,
at which point the ejecta velocity is
$\sim$\,5700\kms.  Provided that $v_t$ is less than this value,
which is generally the case, we
are insensitive to the plateau transition location.
Our simulations then only depend on the outer profile,
that is the two parameters of Equation~\ref{eq:ejecta}:
the normalization of the power-law profile
($C=$\,2.03\tttt{85}~g\,cm$^6$\,s$^{-6}$)\,\footnote{\,The
value and units of $C$ do not mean much at face value; following
many authors we can instead specify the value of $\rho_{\rm ej}$ at
some appropriate reference values of $v$ and $t$.
For $v=$\,1\tttt{4}\kms\ and
$t=$\,10~years we get $\rho_{\rm ej}\approx$~390\amucc, 
very similar to the value of 360\amucc\ used by \citet{BBMcC97let}.}
and the exponent ($n=9$ from \citet{Luo91b}).  
Implicitly fixing these two, we are then left with a degenerate
degree of freedom, our insensitivity to $v_t$, in choosing the ejecta parameters.
Hence there are pairs of ($E_{\rm ej}$,\,$M_{\rm ej}$)
that will equivalently give the hydrodynamics presented here:
(2.4,\,16), (1.5,\,7.9), or (1.0,\,4.3), where the energy is in units
of 10$^{51}$~erg and the mass is in solar masses.  It's important
to note that we do not make any determination of the \textit{actual}
global values of these parameters, they are only used to define the outer ejecta
properties relevant to our hydrodynamics.

%%%%%%%%%%%%%%%%%%%%%%%%%%%%%%%%%%%%%%%%
\subsection{\hii\ Region Parameters}

Estimates of the parameters of the \hii\ region are determined mainly
from observations. This was first done by \citet{Lundqvist99} by
analyzing the ultraviolet line emission from
\sna. \citet{Dwarkadas07RevMex, Dwarkadas07AIP} attempted to refine these
parameters by calculating the radius and velocity of the expanding SN
shock wave in a spherically symmetric case, comparing the results to radio observations,
and iterating until a good fit was obtained. They also calculated a
reasonable fit to the hard X-ray emission under collisional-ionization
equilibrium (CIE) conditions.

In the present work we further develop these earlier
calculations. The \hii\ region radius and density profile are adjusted
so that our simulated emission
agrees with the early X-ray light curves (up to day $\sim$\,5000);
in particular, producing a reasonably sharp ``turn-on'' at $\sim$\,1400 days
and agreeing with the HETG-99 and ACIS-00 measured spectra at later times.
A further constraint on the \hii\ region, especially near the equatorial
plane, comes from the requirement that the FS reaches the
ER (and/or protrusions) at a time {\it and} radius in agreement with
the X-ray imaging observations \citep{Racusin09}.
The parameters we have found appropriate for the \hii\ region,
shown in Figures~\ref{fig:initprofile},
have it beginning at 3.61\tttt{17}\,cm (0.117~pc) with a density of
$\sim$\,130\amucc\ 
and gradually increasing to $\sim$\,250\amucc\ at 6.17\tttt{17}\,cm (0.20~pc)
and remaining at this density to larger radii.
These X-ray-based values have an inner radius closer to the expectations of
\citet{Chevalier95} and differ from the previous \hii\ values
\citep[beginning at 4.3\tttt{17}\,cm with
$\rho\approx 6$\amucc, increasing to $\rho\approx 200$\amucc\ at
6.2\tttt{17}\,cm]{Dwarkadas07AIP} which were based primarilly on the radio
size at early time;
we discuss the reconciliation of these further in \S\,\ref{sec:locs} and
\S\,\ref{sec:img}.

%%%%%%%%%%%%%%%%%%%%%%%%%%%%%%%%%%%%%%%%%%%
\subsection{Equatorial Ring Parameters}
\label{sec:er}

Unlike the \hii\ region where the simple 1D approximation may suffice, it is
clear that only a far cruder approximation to the X-ray emitting ER can be made
with simple 1D constructs.  Note that in this paper we often use the term ``ER'' as a
generic term to refer to the dense ($\sim$\,10$^4$\amucc) equatorial material
with which the SN shock interacts.
The ER therefore includes the finger-like extensions that
\citet{McCray07} has postulated as extending inwards from the ring.
Consequently, the ER is not located at any single radius; this is
clearly demonstrated
by the gradually increasing number of optical hot spots
\citep{Sugerman02}.
A superposition of ER collisions in time is thus expected to make
up the ER contribution to the X-ray light curve.

\begin{figure*}[t]
\begin{center}
\includegraphics[angle=270,scale=0.55]{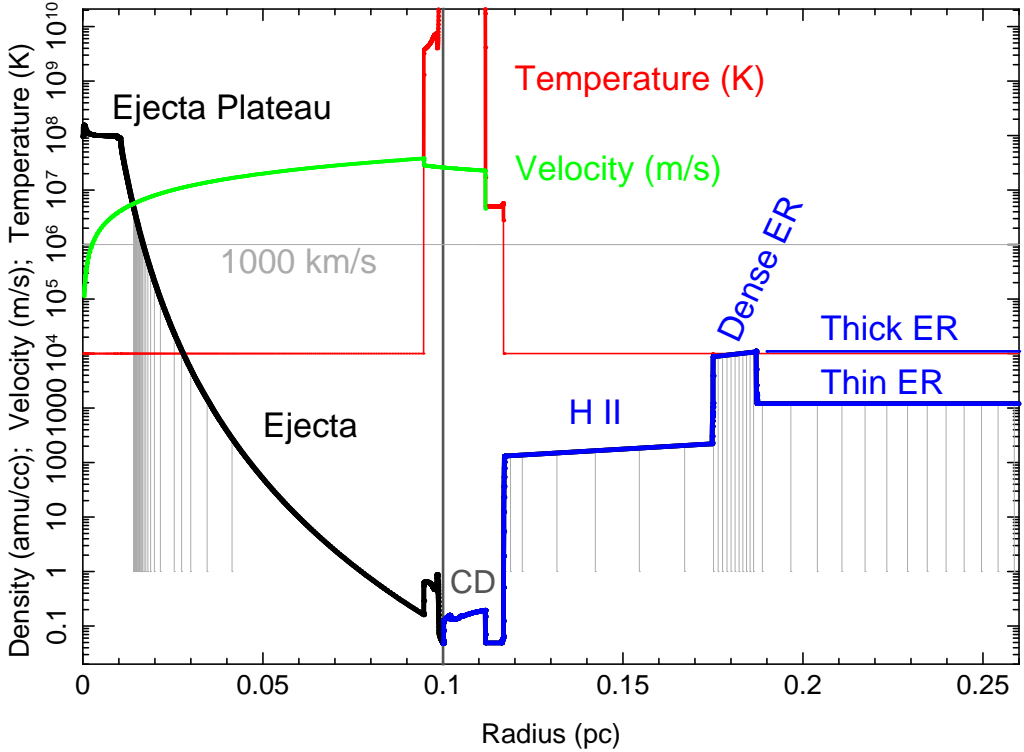} \\
\end{center}
\caption{Initial 1D model radial profiles.
This early-time ($\sim$\,900 d) configuration shows the 
ejecta, the \hii\ region, and the options of a ``thick'' or ``thin'' ER.
The ejecta (black) and CSM (blue)
densities are separated at the contact discontinuity (CD); the
mean plasma temperature (red) and velocity (green) are also plotted.
The gray vertical lines indicate the boundaries of the mass-shells that
are tracked and used to compute the NEI X-ray emission; note that they
do not need to extend into the plateau region of the ejecta profile.
\label{fig:initprofile}}
\end{figure*}

For modeling purposes we choose a single location of the ER starting at
a radius of 5.4\tttt{17}\,cm (with a density of $\sim$\,9000\amucc).
As we'll see, this choice reproduces the ``kink'' in the measured X-ray radius
vs time (\S\,\ref{sec:locs}) 
and the bulk of the dramatic soft X-ray increase after day 6000
(\S\,\ref{sec:lcs}); hence this model component is
representative of the majority of shocked ER material.
Since there is an indication that the X-ray flux
may be leveling off \citep{Park11} we also consider the case of a
finite or ``thin'' ER with a density that then drops from
$\sim$\,10800\amucc\  to (a lower, arbitrary
value of) $\sim$\,1200\amucc\ at a radius of 5.77\tttt{17}\,cm,
as shown in Figure~\ref{fig:initprofile}.

% Show the initial density profile of the with ER hydro
% Plots made with  `Vikram/SN1987A_2001/vs_wER_110802_locFigs.sl

% Show further developed hydro profiles,
% Plots made with  `Vikram/SN1987A_2001/vs_wER_110802_locFigs.sl
\begin{figure*}[t]
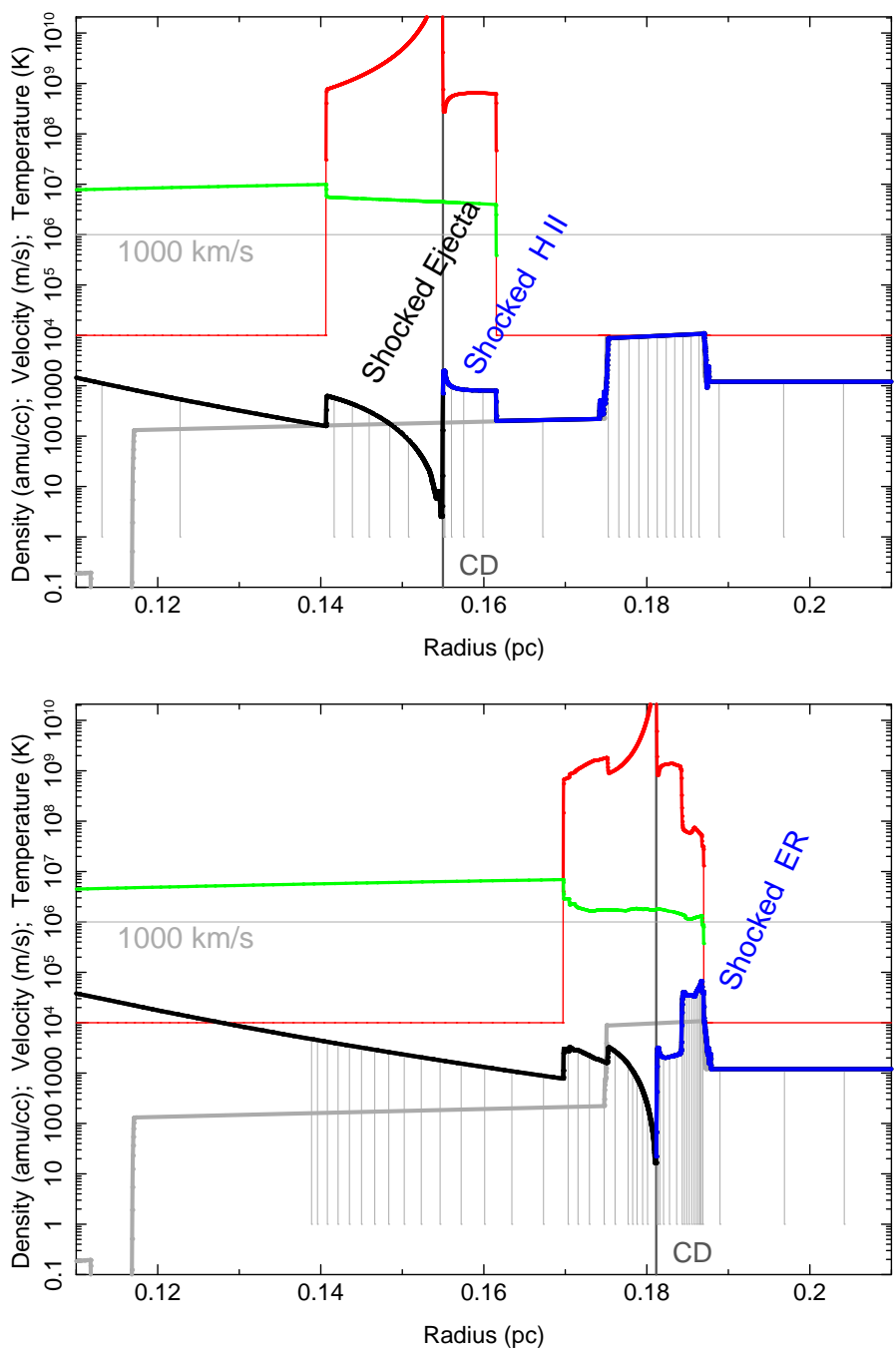

\begin{center}
\includegraphics[angle=270,scale=0.48]{vmov_locs_1060.ps} \\
\vspace{0.1in}
\includegraphics[angle=270,scale=0.48]{vmov_locs_1103.ps} \\
\end{center}
\caption{Model profiles at later times.
{\it Top}: The simulation at the ACIS-00 epoch when the FS
has moved well into the \hii\ region; note that the early-time
density profile is shown in gray for reference.
{\it Bottom}: At the HETG-11 epoch, the FS has encountered
the ``thin ER'', slowed, and is just about to exit the ER.
The color coding is the same as for Figure~\ref{fig:initprofile}.
\label{fig:modprofile}}
\end{figure*}

\clearpage
%%%%%%%%%%%%%%%%%%%%%%%%%%%%%%%%%%%%%%%%
\section{MODEL RESULTS AND COMPARISONS}
\label{sec:results}

The previous section summarized the ejecta-CSM properties of our
hydrodynamic simulations.  In the following we compare several results derived
from the simulations with the observed properties of \sna.

\subsection{Model Locations}
\label{sec:locs}

% Show hydro radii vs measured radii (radio, X-ray)
% Created by hylocs.sl in `isis_110101/, renamed the .ps output.
\begin{figure*}[t]
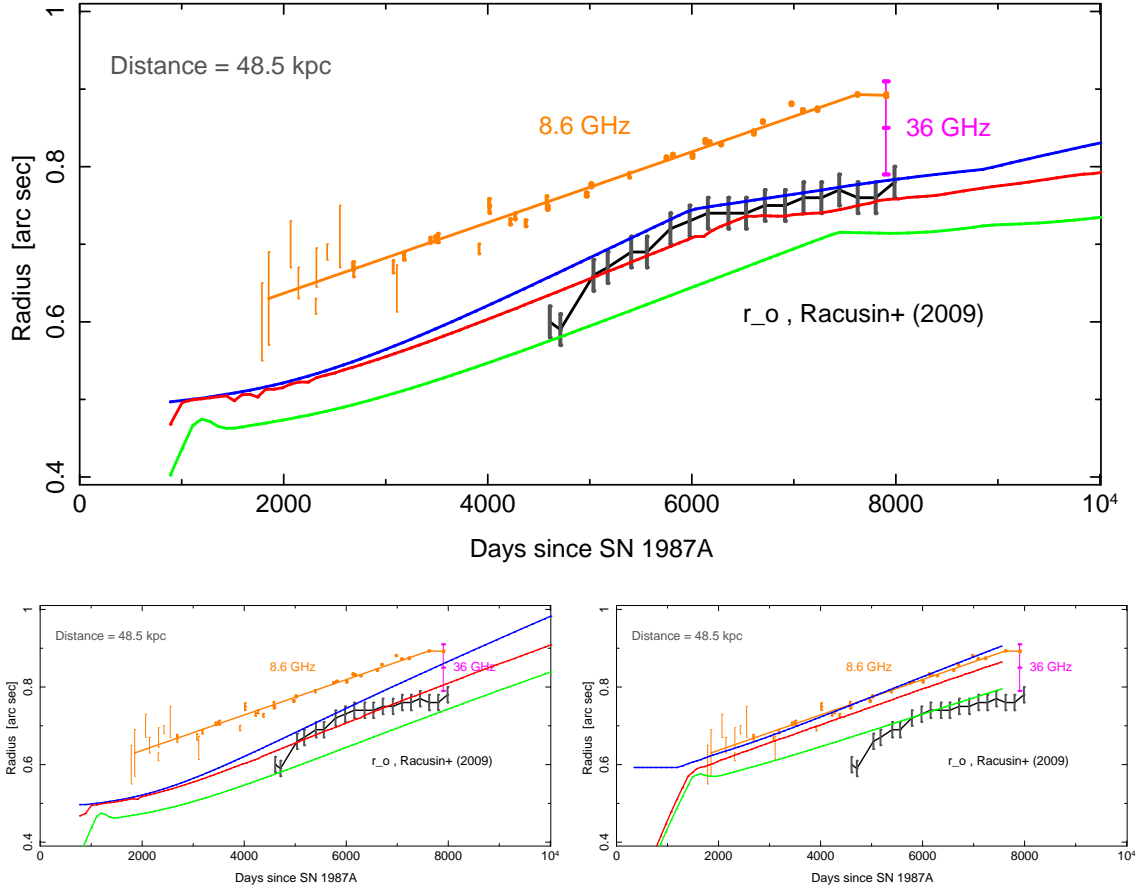

\begin{center}
\includegraphics[angle=270,scale=0.60]{sn87a_wER_110802_locs485.eps} \\
\vspace{0.2in}
\includegraphics[angle=270,scale=0.30]{sn87a_noER_110718_locs485.eps} 
\includegraphics[angle=270,scale=0.30]{sn87a_noER_090410_locs485.eps} \\
\end{center}
\caption{Model radii compared with X-ray and radio measurements. 
  {\it Top}: The ``with ER'' simulation has most X-ray emission between the FS (blue)
  and CD (red) radii and their locations
  agree very well with the ACIS-measured average radii (black).
  Note how the kink due to the ER collision occurs first in the FS curve
  (around day 6000) and then at later times in the CD and RS (green) curves.
  {\it Lower-left}: As expected, the ``no ER'' simulation shows
  continued expansion; this applies to the case of the out-of-plane \hii\ region
  giving rise to the very-broad X-ray component.
  {\it Lower-right}: The 8.6~GHz average radii (orange) are better
  matched with a different CSM profile, here we use the one given in
  \citet{Dwarkadas07AIP}.
\label{fig:modradii}}
\end{figure*}

The radio and X-ray measured radii are used to constrain the
hydrodynamic models.  In Figure~\ref{fig:modradii} we show these radii
along with the locations of the simulations' FS, CD, and RS.
The measured radii are given in arc seconds and are based on
de-projected model fitting for both the radio
\citep{Ng08} and X-ray \citep{Racusin09} values.
To convert from the simulation spatial units (cm) to observed angular units
a distance of 48.5~kpc has been used to get the agreement shown;
this is very close to the accepted distance to \sna\ 
(\S\,\ref{sec:intro}) and further ``fine tuning'' of the hydrodynamics
is not warranted for these simple models.

For the ``with ER'' simulation, Figure~\ref{fig:modradii} top panel,
the locations of the model emitting region (between the FS and
CD) are in good agreement with the measured ACIS radii and reproduce
the ``kink'' that occurs at about day 6000
at a radius of $\sim$\,0.74 arc seconds \citep{Racusin09}.
This radius is set by the start of our ER at 5.4\tttt{17}~cm, or 0.744 arc
seconds, and it
also agrees well with the inner radii of the majority of the optical ring
material that is seen in the very useful Figure\,7(b) of \citet{Sugerman02}.
Hence, our ER model location is a good representation of the 
location of the start of the bulk of the ER.
Regarding a possible end to the ER, our ``thin ER'' case
ends at a radius of 5.77\tttt{17}~cm, or 0.795 arc seconds.
This gives our thin ER a radial extent of $\sim$\,0.05 arc seconds, and,
looking again at Sugerman's Figure 7(b), we see that this is
a reasonable radial extent for the central portions
of {\it individual} bright optical features.
At late times,
the locations shown in Figure~\ref{fig:modradii} are
based on the ``thin ER'' profile (labeled in Figure\,\ref{fig:initprofile})
which has the FS leaving the
dense ER at $\sim$\,9000 days; this
produces a slight upward kink in the FS location curve at this time.

The ``no ER'' simulation, representing the out-of-plane \hii\ region,
does a good job of matching the ACIS radii before day 6000
(since it is the same as the ``with ER'' simulation at those times)
and continues to expand at a roughly constant rate.
By 25 years, the FS in the \hii\ region simulation has progressed to
a distance of $\sim$\,6.8\tttt{17}\,cm (0.22~pc),
equivalent to $\sim$\,0.94 arc seconds for the model-scaling distance of 48.5~kpc.
Comparing this to the mean radius and extent of the optical ER,
0.83 and 0.65--1.0 arc seconds, respectively
\citep[Figure\,7]{Sugerman02},
suggests that the FS in the out-of-plane \hii\ region is now beyond
most of the ER.

Comparing the model locations with radio measurements on these plots,
the 8.6~GHz data \citep{Ng08}
appear to be different from both the ``with ER'' and
``no ER'' X-ray-based models; in particular showing a $>0.6$~arc second
radius by day 2000.
However, using the previous CSM profile of \citet{Dwarkadas07AIP}
does give a CD--FS location curve that is in good agreement 
with the measured average radii for the 8.6\,GHz radiation,
Figure~\ref{fig:modradii} lower-right panel.
The single 36\,GHz radius measurement \citep{Potter09} is located
between the X-ray and the 8.6~GHz values, although it has an
uncertainty range including each of these.
One way to accomodate the larger 8.6~GHz radio radii is to have that
emission come from material further from the equatorial plane as
schematically shown in Figure~\ref{fig:hydrogeom}.

%%%%\clearpage
%%%%%%%%%%%%%%%%%%%%%%%%%%%%%%%%%%%%%%%%
\subsection{Model Light Curves}
\label{sec:lcs}

% Show X-ray light curves from hydro and data
% Created by hylc.sl in `isis_120201/
\begin{figure*}[t]
\begin{center}
\includegraphics[angle=270,scale=0.65]{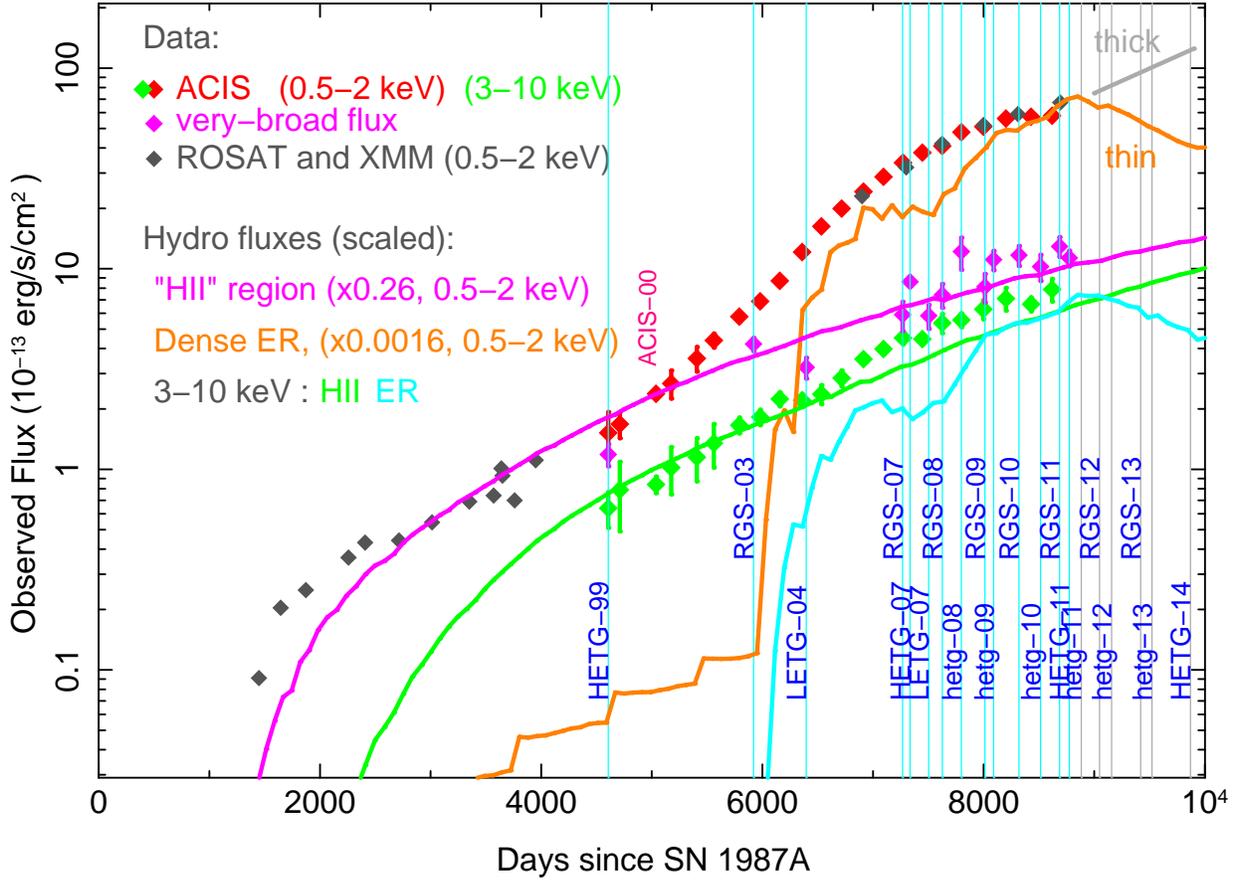} \\
\end{center}
\caption{ Comparing X-ray light curves with the hydrodynamic model fluxes.
The \hii\ region emission from the ``no ER'' model
is shown by the curves beginning before day 3000 (magenta, green).
The model does a reasonable job of matching the initial and late-time 3--10~keV flux
(green diamonds)
as well as modeling the initial 0.5--2~keV turn-on and growth (ROSAT points). 
At late-times the 0.5--2~keV flux from the \hii\ region  
is in reasonable agreement with the measured very-broad flux (magenta
diamonds), strengthening the identification of the \vb\ emission
with the \hii\ region.
The emission from the shock collision with the dense ER 
rises steeply around day 6000 (orange, light blue).
The light curve here (orange) is for the
the ``thin (or finite) ER'' case
of Figure\,\ref{fig:initprofile}; the gray line indicates the
0.5--2~keV flux when the ER continues in the ``thick ER'' case.
\label{fig:modlc}}
\end{figure*}

We use the hydrodynamic simulations as input to calculate the X-ray emission from each
shocked mass-cut shell at each time step, employing the same scheme
to track the NEI plasma state and carry out spectral calculations
as outlined in \citet{Dwarkadas10}.
Once the hydrodynamic solution is determined -- fixing temperatures,
ionization ages, and normalizations for the mass-cuts -- the only free parameters are the
abundances,
an optional clumping set by a clumping fraction and its
over-density, and the overall normalization set by the solid angle
which each 1D model actually fills, $\Omega_\shii$ and $\Omega_{\rm ER}$.
The abundances can be determined by fitting to the measured spectra,
as described further in \S\,\ref{sec:hiiabunds}\,\&\,\ref{sec:erabunds}.
In terms of line broadening, the hydrodynamic velocities of the shells
are used to include Doppler broadening in the synthesized spectra;
in \S\,\ref{sec:vbspectrum}
this ``real'' broadening is compared with the simple smoothing
approximation that was used to quantify the \vb\ fraction
(\S\,\ref{sec:vb}).
Finally, for high shock velocities and low $\tau$'s, the simple $\propto\,v^{-2}$
behavior of $\beta=T_e/T_i$ \citep{Ghavamian07} produces $T_e$ values that are very low.
As in \citet{Dwarkadas10} we have used a
modified $\beta(v)$ relation, suggested by the results of
\citet{vanAdelsberg08}, to
set a minimum value of $\beta_{\rm min}\approx 0.06$.

Integrating the simulated spectra over the standard energy bands,
0.5--2~keV and 3--10~keV, we created model light curves, shown in
Figure~\ref{fig:modlc}.  The light curves for the  two hydrodynamic simulations
have been plotted separately and their (imagined) sum can reasonably describe
the observed \sna\ light curves over the full time span.

\clearpage

A filling fraction of $\Omega_\shii/4\pi=$~0.26 is
used to match the overall flux level for the \hii\ (``no ER'')
simulation, corresponding to emission from
$\pm$\,15 degrees of the ring plane.  This is roughly similar to 
other estimates of the \hii\ region's out-of-plane extent, e.g., 
by \citet[$\pm$\,30 degrees]{Michael03},
\citet[$\pm$\,10 degrees]{Zhekov10}, and \citet[$\pm$\,7.2 degrees]{Mattila10}.
The abundances for the \hii\ light curve are set from fits
to the early ACIS observation, described in \S\,\ref{sec:hiiabunds}.
One characteristic of this early-time \hii\ emission is
its relative hardness, e.g., for the deep ACIS-00 observation
at day 5036 the ratio $F_{3-10} / F_{0.5-2}$ is 0.35
whereas at later times (day 7997, January 2009) the ratio has
dropped to 0.12.  As the light curves show, this is caused by the
increase in the 0.5--2~keV flux due to the collision with
the ER, rather than any decrease in the 3--10~keV emission.

In contrast to the \hii\ region, the ER needs to produce
emission primarily in the 0.5--2~keV range while keeping
the 3--10~keV flux below the measured levels.
Even though the ER is dense, our simulation still gives
electron temperatures at/above 3~keV and
thus relatively too much 3--10~keV flux.
We could further increase the density of the ER, however given the
resonable agreement of the slope of the ``with ER'' simulation
(top panel of Figure~\ref{fig:modradii}) and the expectations of an inhomogeneous
ER (e.g., the optical spots), we consider another, approximate approach.
We can match the observed spectra
by including emission from ``clumps'' in the ER
with a density enhancement $\eta_{\rm clump}\approx \times$5.5 that
fills $f_{\rm clump}\approx$ 30\% of the CSM volume.
The clumping is implemented at the spectral synthesis stage and
introduces $kT_e\approx$ 0.5--0.8~keV emission components
greatly enhancing the 0.5--2~keV emission (further details of the
clumping implementation are given in \S\,\ref{sec:erabunds}.)
We acknowledge that for these clumping parameters most of the mass is in the very dense
clumps (i.e., 0.30$\times$5.5\,/\,(0.70$\times$1+0.30$\times$5.5)
~$\approx$ 70\%)
and so the clumping is not a small perturbation on the
ER hydrodynamics; therefore our 1D clumped ER simulation is just a placeholder
for a realistic multi-dimensional simulation.

The single set of ``with ER'' light curves plotted in Figure~\ref{fig:modlc}
represents the bulk of the 0.5--2~keV flux increase; the fractional area used to
scale the 1D model's emission is
very small: $\Omega_{\rm ER}/4\pi =$~0.0016 .
To put this into perspective,
the fraction subtended by a uniform region with the dimensions of
the optical ring, i.e., a region extending
$\pm$\,0.05 arc seconds out of the plane at a radius of 0.83 arc seconds,
is almost 40$\times$ larger, $\sim$\,0.0600.
Hence, the small value indicates that the the ring is not uniformly
dense and clumped, but rather it has ``X-ray hot spots'' analogous to
the (still) discrete optical spots around the ring.
Specifically, the 0.0016 value would arise from $\sim$\,20 regions
each subtending a diameter of $\sim$\,2 degrees around the ring.

It is seen in Figure~\ref{fig:modlc} that there is 
earlier-time flux growth above the \hii\ emission that our plotted ER
curve does not include.  As mentioned in \S\,\ref{sec:er}, a range of inner radii is
expected for the ``ER'' with some material at radii smaller than the
0.74~arc seconds where our modeled ER starts.
Mentally ``translating'' our ER curve, we would expect to get
flux beginning $\sim$\,5500 days
by having a small fraction of ER material (perhaps
0.1\,$\times$\,$\Omega_{\rm ER}$) that is impacted
at $\sim$\,4600 days.  From our hydrodynamics this time
corresponds to a radius of 4.8\,\tttt{17}\,cm (0.155~pc)
or 0.67 arc seconds.  This radius is
in reasonable agreement with the locations of the early optical spots
\citep[Figure\,7]{Sugerman02}.

%%\clearpage
%%%%%%%%%%%%%%%%
\subsection{\hii\ Abundances from the Hydrodynamics}
\label{sec:hiiabunds}

Using the set of shocked mass-cut shells at a given time-step of the 
hydrodynamics, we can create a spectral model in ISIS which is the sum
of emission from each of the shells.  The relative normalizations, temperatures,
and ionizations ages for each model component (mass-cut shell) are set from the
hydrodynamics.  Tying the abundances of all components together,
we then have a model which can be fit to measured spectra
by adjusting the abundances and an overall normalization.
By using the hydrodynamics to set the plasma properties we reduce the
degeneracy in fitting that arises when the temperature(s) and ionization
age(s) of the plasma are unconstrained and therefore we get accurate abundance
values.

% Temp-Tau plots of hydro at Obs 1967 epoch: noER_110718 step 60
% created with plot_Ttaudist.sl in `Vikram/SN1987A_2011/
%   and
% Obs 1967 data with hydro-based model, from isis_110101/obs_1967_compare.sl
\begin{figure*}[t]
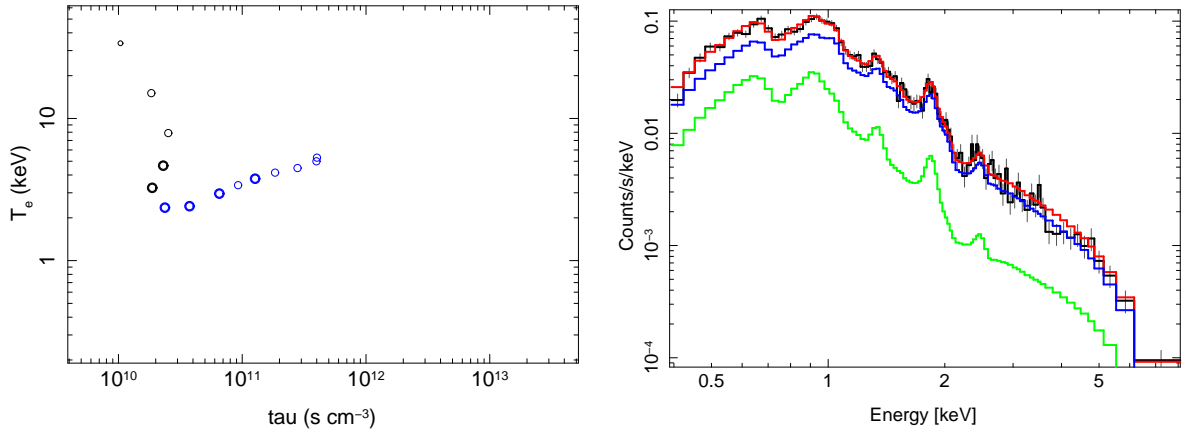

\begin{center}
\includegraphics[angle=270,scale=0.30]{110718_Ttaudist_1060.ps}
\hspace{0.1in}
\includegraphics[angle=270,scale=0.30]{obs1967_hydro-based.ps} 
\end{center}
\caption{{\it Left}:  The distribution of $kT_e$ and $\tau$ values in
the \hii\ hydrodynamics at the ACIS-00 epoch.
Points of $(\tau, kT_e)$ are plotted for all shocked
mass shells in the model;
symbols for the shocked ejecta (black)
and CSM (blue) are larger/bold for brighter shells.
{\it Right}: The ACIS-00 data (black) are fit with a model (red) that
is the sum of the emission from the 14 shocked shells of the left panel;
a single norm and the abundances of N, O, Ne, Mg, Si, S, and Fe were the
only free parameters in the fit.  The emission from the CSM shells (blue)
is much larger than that of the shocked ejecta (green) at this early epoch.
\label{fig:TTauHII}\label{fig:obs1967}}
\end{figure*}

\input{tab_obs1967}

The early deep \chandra\ observation, ACIS-00, is the best data set to use
to determine the abundances of the \hii\ region (as opposed to the
dense ER abundances, which may differ).
These data also provide a useful example of how the fit abundances and
their confidence ranges vary
depending on the $kT$--$\tau$ model assumptions.  The results of fitting four
different cases are shown in Table~\ref{tab:obs1967}.
In all cases the redshifts and $N_H$ were frozen
and the normalization(s) and abundances (N, O, Ne, Mg, Si, S, and Fe) were fit.
In the first case
a single {\tt vpshock} component is fit giving $kT\approx 3.0$~keV
and $\tau\approx$ 1.0\tttt{11}\utau; these are similar to the values in \citet{Park02}.
As a second case, we fit a 2-shock model fixing the temperatures,
ionization ages, and $N_H$ based on values from \citet{Zhekov10};
this has a high temperature component similar to the 1-shock value
along with a very low temperature component.
In the third case, our model of Table~\ref{tab:3shock} is shown, and
because we have set $N_{\rm mid}$ to 0 for this early data it is
also a 2-shock model with a different set of $kT$'s and $\tau$'s.
Finally, we fit the data with the temperatures and ionization ages
fixed based on the 
values from the ``no ER'' hydrodynamic model at the ACIS-00 epoch;
these values are plotted in the
left panel of Figure~\ref{fig:TTauHII} and correspond to the model
at the epoch shown in the top panel of Figure~\ref{fig:modprofile}.

The reduced $\chi^2$ values are near 1.0 for all four of the fits, so they
are equally acceptable from a data-fitting perspective.  However, as the table
shows, the 1-$\sigma$ ranges of the fit abundances depend very much on the
$kT$--$\tau$ values used in the model.  
If it is possible to constrain the $kT$--$\tau$ values through
other means or assumptions, as in the fourth case here, then
the fit abundances will be both more realistic and better constrained.
Hence, we adopt the hydrodynamics-based model as the most physical and
include its abundances in the \hii\ column of Table~\ref{tab:abundsdisc}.

\input{tab_abundsdisc}

\clearpage
%%%%%%%%%%%%%%%%
\subsection{ER Abundances and Clumping}\label{sec:erabunds}

% Temp-Tau plots of hydro at HETG-11 epoch
%  noER_110718 step 101
%   wER_110719 step 103
% created with plot_Ttaudist.sl in `Vikram/SN1987A_2011/
%
\begin{figure*}[t]
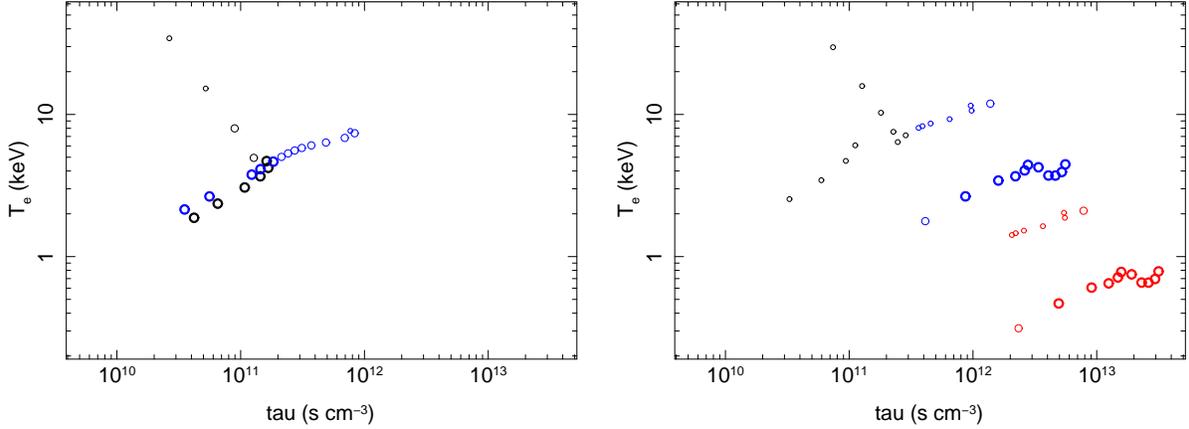

\begin{center}
\includegraphics[angle=270,scale=0.30]{110718_Ttaudist_1101.ps}
\hspace{0.1in}
\includegraphics[angle=270,scale=0.30]{110719_Ttaudist_1103clmp.ps}
\end{center}
\caption{Distributions of $kT_e$--$\tau$ for the hydrodynamics at the HETG-11 epoch.
Points of $(\tau, kT_e)$ are plotted for all shocked
mass shells in the \hii\ (no-ER, left panel) and ``with ER'' (right
panel) simulations.
Symbols for the shocked ejecta (black)
and CSM (blue) are larger/bold for brighter shells.
The \hii\ distribution is similar to earlier times
(Figure~\ref{fig:TTauHII},
left panel) although shifted to somewhat higher $\tau$ and $kT_e$ values.
The ``with ER'' distribution (right panel) is dominated by components at
much higher $\tau$ values;
the addition of a clumped component introduces additional
$kT_e$--$\tau$ values (red) that are scaled from the
non-clumped CSM values.
\label{fig:TTauHETG11}}
\end{figure*}

We can similarly fit hydrodynamics-based models to the later
observations where the emission is dominated by the shocked ER material
and in this way determine the ER abundances.
At these times there will still be a contribution from the \hii\
region, i.e., the \vb\ component.  This contribution is removed
from the data before fitting by evaluating our ``no ER'' model at the
appropriate epoch and subtracting the predicted counts from the data
before proceeding with ER-model fitting.

The $kT_e$--$\tau$ values of the shocked mass-cuts in our \hii\ and
ER models at the HETG-11 epoch are shown in Figure~\ref{fig:TTauHETG11}.
As the blue points in the ``with ER'' (right) panel show, the $kT_e$
values in the shocked dense ring are mostly at/above 3~keV.  Given
these temperatures, simply adjusting the abundances is insufficient to
fit the data, there is simply too little low-$kT_e$, high-$\tau$ plasma.  
We have therefore included ``clumps'' in the ER material
following the scheme in \citet{Dwarkadas10}.

For each of the shocked CSM mass-cuts (blue circles in the right panel
of Figure~\ref{fig:TTauHETG11})
we add an additional model component 
with its temperature reduced by the clump's density enhancement,
$kT_e^\prime = kT_e/\eta_{\rm clump}$, and its
ionization age increased to $\tau^\prime = \eta_{\rm clump}\,\tau$.
The additional clumped components are shown as
red circles in the right panel of Figure~\ref{fig:TTauHETG11}
for $\eta_{\rm clump}\approx$ 5.5.
The original and clumped components have additional
factors in their norms of $(1-f_{\rm clump})$ and
$f_{\rm clump}\,\eta_{\rm clump}^2$, respectively,
where $f_{\rm clump}\approx$ 30\% is the volume fraction
of clumped material.
The resulting ISIS fit function at the HETG-11 epoch consists of 45
NEI model components but with only 9 free parameters: an overall 
normalization, the 2 clumping parameters, and the ER abundances of O, Ne,
Mg, Si, S, and Fe.

% Hydro-based spectral components at 2007 and 2011 epochs.
% From plot_data_hydro.sl in isis_120201/.
% Side-by-side for size of 0.32.
\begin{figure*}[t]
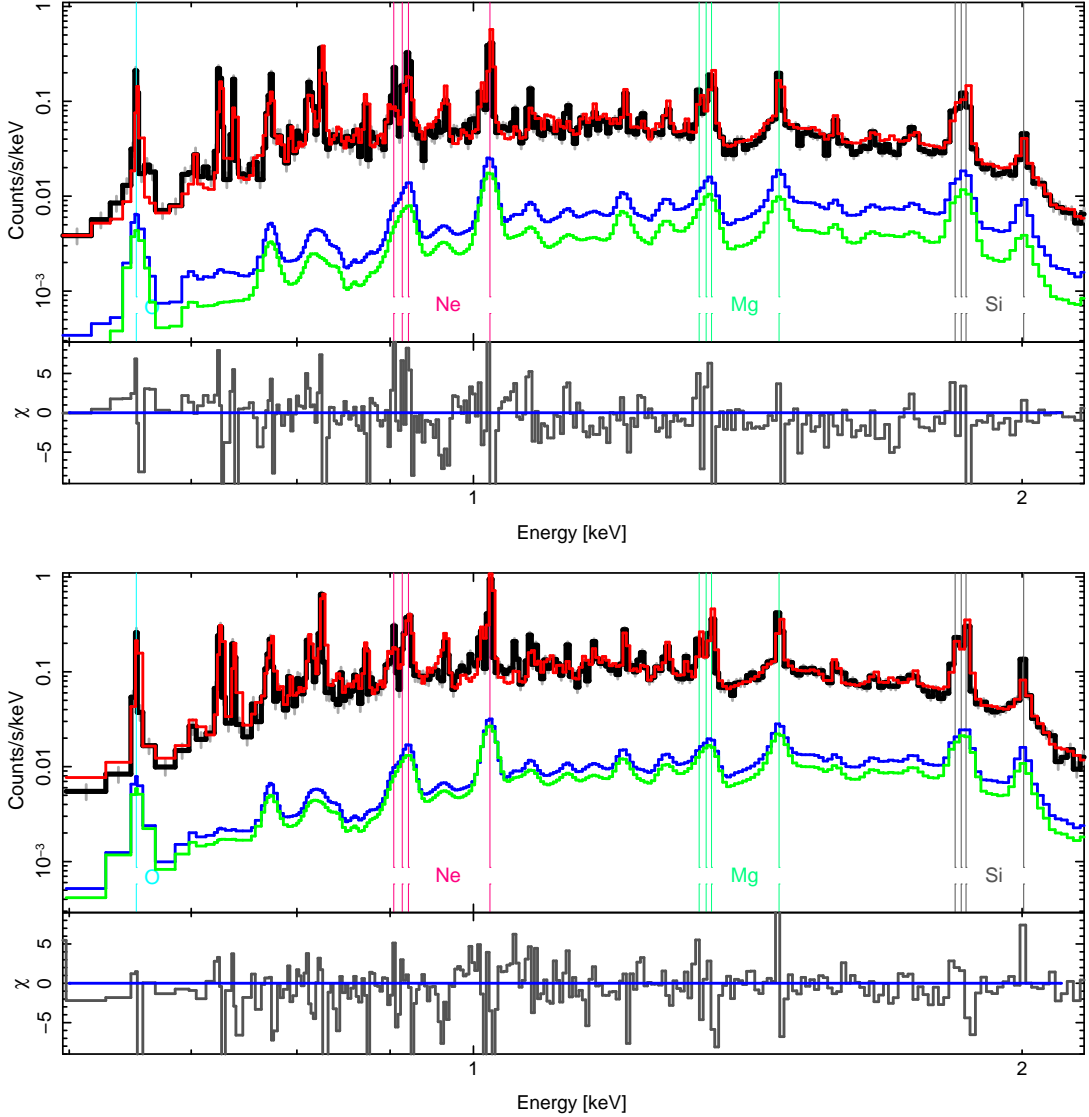

\begin{center}
\includegraphics[angle=270,scale=0.60]{HETG-07_data_hydro_MEG.ps} \\
\vspace{0.1in}
\includegraphics[angle=270,scale=0.60]{HETG-11_data_hydro_MEG.ps}
\end{center}
\caption{HETG(MEG) spectra and full hydrodynamics-based models.
The HETG-07 (top) and HETG-11 (bottom) data (black) are reasonably
fit by the total (\hii\ plus nominal clumped-ER) model spectra (red).
The \hii-shocked-CSM (blue) and the \hii-shocked-ejecta (green)
components are also shown individually.
Note the relative growth of the shocked ejecta (green) between the epochs;
hence at late times the ejecta and its abundances will dominate the
\hii\ (i.e., \vb) component.
\label{fig:hydrospect}}
\end{figure*}

We fit both the deep HETG-07 and HETG-11 data sets with their appropriate-epoch
hydrodynamics-based models, determining the abundances and clumping
parameters for each.  There are only small differences
between the 2007 and 2011 fits,
so we have adopted nominal ER parameters that are the average values given in the
``ER'' column of Table~\ref{tab:abundsdisc}.
Note that the N abundance was set based on similar fitting
to the RGS-11 spectrum.
These values were then used
to generate our light curve (\S\,\ref{sec:lcs})
and to fit the hydrodynamics-based models to all of the grating spectra
with only a single normalization adjustment.  As examples,
Figure~\ref{fig:hydrospect} compares the two deep HETG spectra
with the sum of the hydrodynamics-based emission from the \hii\ region and
the nominal-parameter ER emission.  

Given our crude clumping implementation and the system's physical
complexity, we have
less confidence in the accuracy, i.e., agreement with reality, of our ER abundances
than those we determined for the \hii\ region at the ACIS-00 epoch.
However since the ER $kT$s and $\tau$s are based on a hydrodynamic
solution there is some expectation that they will be an improvement over
simpler multi-shock models.
For comparison, we've also included in Table~\ref{tab:abundsdisc}
the abundances given by
\citet[multi-shock fits to grating data]{Zhekov09},
\citet[RSS fits to the ACIS data]{Zhekov10}, and
\citet[optical \& NIR data before the ER collision]{Mattila10}.
Our ER abundances for N, O, Ne, and S do appear to be in someshat
better agreement with the \citet{Mattila10} values
than are the other X-ray-determined abundances.

This is a good place to compare the free parameters of the
\citet{Zhekov10} model (Z10 in the following)
with ours, in terms of their number and their values.  
The two model geometries are similar at later times, both are dominated by the
ER emission, Figure~\ref{fig:rsscompare}, and both make approximations
to include clumped ER material.
Each model fixes time-independent values for the CSM's $\rho(r)$,
the $N_H$, and the abundances.
In addition to these, Z10 fit six parameters
{\it for each observation}: the values
of the temperature, ionization age, and emission measure for both the 
blast wave (BW) and the transmitted shock (TS).
For example, for their latest data point which is near the HETG-07
epoch, the values for these are: $kT_{\rm BW}\approx$\,1.7~keV, 
$kT_{\rm TS}\approx$\,0.33 keV,
$\tau_{\rm BW}\approx$\,1.4\tttt{11}\utau,
$\tau_{\rm BW}\approx$\,30\tttt{11}\utau, and the respective 
emission measures are 1.05 and 9.5 in units of 10$^{59}$\,cm$^{-3}$.
In contrast, all of our mass-shell $kT$ and $\tau$
values are fixed by the hydrodynamics, e.g., as for HETG-11
in Figure~\ref{fig:TTauHETG11}, and our ER modeling adds only three
further time-independent parameters: the overall normalization,
$\Omega_{\rm ER}$, and the two clumping parameters.
In our approximation the clump over-density, $\eta_{\rm clump}\approx
5.5$, sets the ratio of ER-to-clump temperature.
For the Z10 values above, this
ratio is $T_{\rm BW}/T_{\rm TS}\approx 5$, close to ours.
Likewise, the Z10 emission measure ratio, ${\rm EM}_{\rm TS}/{\rm EM}_{\rm BW}\approx 9$,
is in the ball park
of our value which is given by $f_{\rm clump}\eta_{\rm clump}^2/(1-f_{\rm
clump})\approx13$, using our clump volume fraction $f_{\rm clump}\approx 30$\,\%.
In constrast, the clump volume fraction for the Z10 model is of order 5\,\% because
the clump-to-ambient density ratio is large: 
$\rho_{\rm TS}/\rho_{\rm BW}\approx 14$; this latter value is
a function of the temperature ratio as shown in a
plot in the Appendix of \citet{Zhekov09}. 
These comparisons suggest that there may be some degeneracy between over-density
and volume fraction for the clumps.  In anycase, it is clear that both
approaches here are only approximating the complex 2D/3D density structures
of the real ER.

\clearpage
%%%%%%%%%%%%%%%
\subsection{Very-Broad Spectrum from the Simulation}
\label{sec:vbspectrum}

\begin{figure*}[t]
\begin{center}
\includegraphics[angle=270,scale=0.60]{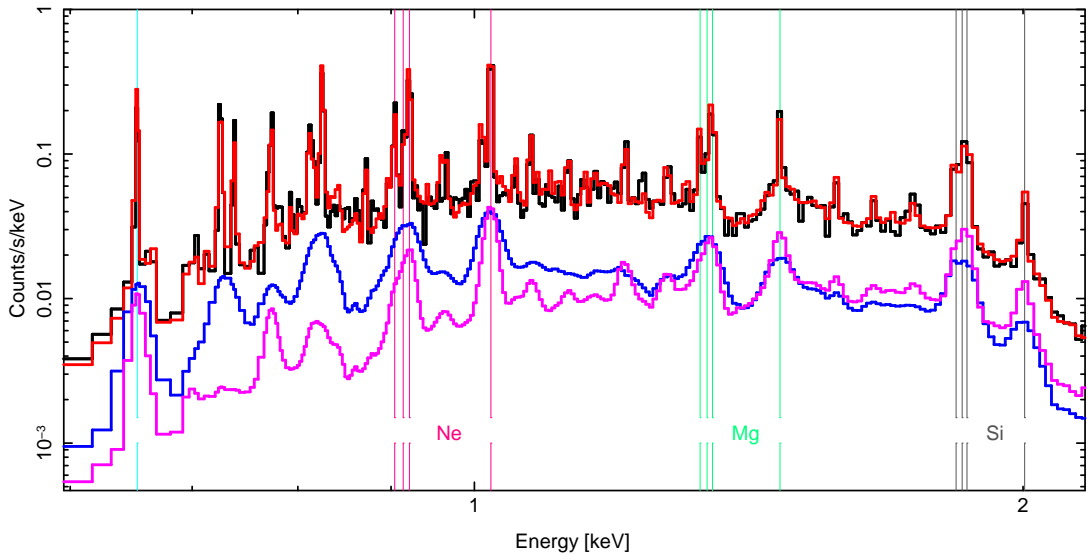} \\
\end{center}
\caption{Comparing the heuristic very-broad spectrum
with the hydrodynamics-based \hii\ \vb\ spectrum.
The HETG-07 data (black) are shown with the best-fit 3-shock
model (red) that includes a \vb\ component (blue).
This ``smoothed version of the model'' form of the \vb\ component is compared
with the multi-shells model spectrum at the HETG-07 epoch
(magenta).  Ignoring the overall continuum shape,
they are in reasonable line-broadening agreement except for 
the lack of significant \fexvii\ lines between 0.70--0.75~keV in
the hydro-based model.
\label{fig:vbhydro}}
\end{figure*}

We can check our decision to use
a smoothed version of the spectrum as the model for the
\vb\ component by directly comparing the two of them, as shown in
Figure~\ref{fig:vbhydro}.  Here, the broad lines in the ``real''
(from the hydrodynamics) \vb\ component are similar to those
in the smoothed version of the 3-shock fit to the data.
This similarity can be roughly explained using the $kT_e$--$\tau$
distributions shown in Figure~\ref{fig:TTauHETG11}: the \hii\ (\vb)
emission is dominated by components with $kT_e \approx$ 2.5~keV
and $\tau \approx$ 6\tttt{10}\utau, whereas the main ER emission
is from clumped material with $kT_e \approx$ 0.7~keV
and $\tau \approx $ 2\tttt{13}\utau.  Putting these values into a {\tt vnei}
model gives similar line structures for the two cases, hence,
the high-$kT$-low-$\tau$ plasma
produces ionization states and emission lines that are similar
to those of a low-$kT$-high-$\tau$ plasma.
Based on this comparison we did, however, decide to ignore the
\fexvii\ range 0.70--0.75~keV when doing the \vb\ fitting as these
lines were clearly not present in the hydrodynamic-based \vb\ spectrum.

% Comparing the very-broad model with the hydro-HII model
% From isis_120201/compare_3shock_noER.sl

In terms of the assumption of a constant width of the \vb\ emission
in time (the evolution path shown in Figure~\ref{fig:vbcontours}),
the average bulk motion of the shocked \hii\ material
in the ``no ER'' simulation varies (only) from 4200\kms\ to
3850\kms\ from the HETG-99 epoch to an age of over 25 years (9000+ days).
This is also seen in
the locations of the ``no ER'' CD and FS locations, lower-left panel 
of Figure~\ref{fig:modradii}, which
follow nearly straight lines with overall slopes of
4300\kms\ and 5100\kms, respectively.

%%\clearpage
%%%%%%%%%%%%%%%%
\subsection{The Mass of the X-Ray Emitting Material}\label{sec:mass}

We can calculate the shocked mass for different components and epochs of the
model.  For the \hii\ region, at the ACIS-00 epoch the
FS is at $\sim$\,0.162~pc (top panel of Figure~\ref{fig:modprofile})
and the mass of shocked CSM in the simulation, correcting for the opening angle
normalization $\Omega_\shii/4\pi\approx$ 0.26 (\S\,\ref{sec:lcs}),
is 0.012 solar masses.  At this time our model shows $\sim$\,0.005 solar masses of ejecta
are shocked within the same opening angle.
Later, at the HETG-11 epoch, 
the shocked \hii\ material that produces the \vb\ emission in our model
has grown to 0.042 solar masses along
with an additional 0.024 solar masses of shocked ejecta.
Our \hii\ mass values compare in order of magnitude
with the mass of 0.018 solar masses that \citet{Mattila10}
require in their low-density (10$^2$ atoms\,cm$^{-3}$) \hii\ component.

The total ``4$\pi$'' mass of the thin ER is 1.18 solar masses;
however, correcting for the small solid-angle fraction,
$\Omega_{\rm ER}/4\pi\,\approx\,0.0016$,
we get only $\sim$\,0.0019 solar masses of ER that are shocked at the
HETG-11 epoch (when the FS is exiting the thin ER).
Including the extra mass due to clumping (multiplying by a factor of 2.35, \S\,\ref{sec:lcs}),
we have 0.0045 solar masses as the total shocked-ER mass that produces the observed
non-\vb\ X-rays in our model at the HETG-11 epoch.
This is to be compared with an estimate of the total mass of the UV-ionized ER
of 0.058  solar masses distributed among three densities of 1.4\tttt{3},
4.2\tttt{3}, and 4.2\tttt{4}\amucc\ \citep[using
$\sim$\,1.4~amu per atom]{Mattila10};
of this total, 0.0120 solar masses are in the highest-density component,
almost 3 times our X-ray inferred mass.
However, it could well be that our ER X-ray emission is a factor of a
few overly efficient per mass and including this as an error bar puts us
in a gray area in trying to use this mass comparison
to decide if all of the ionization-visible dense ER has been shocked at the
current $\sim$\,25 year age of \sna.

%%%%%%%%%%%%%%%%
\subsection{Image Implications}
\label{sec:img}

% Hydro-based geometry from make_sequence.sl in e2d_110322/
\begin{figure*}[t]
\begin{center}
\includegraphics[angle=0,scale=1.0]{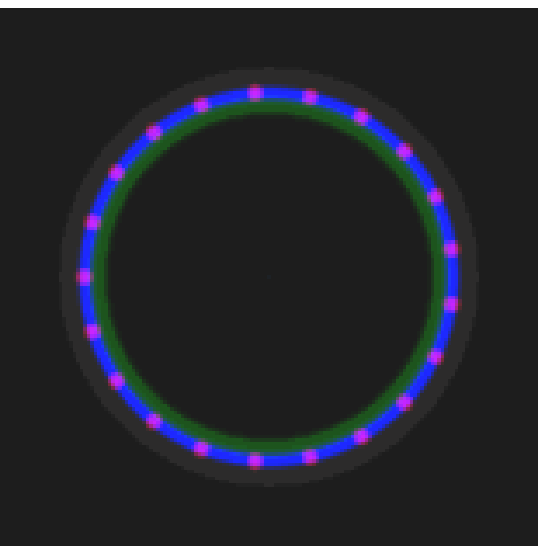}
\hspace{0.15in}
\includegraphics[angle=0,scale=1.00]{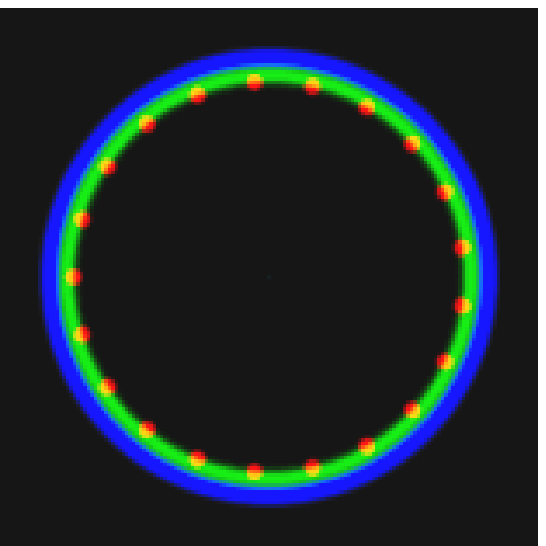} \\
\vspace{0.05in}
\includegraphics[angle=0,scale=1.0]{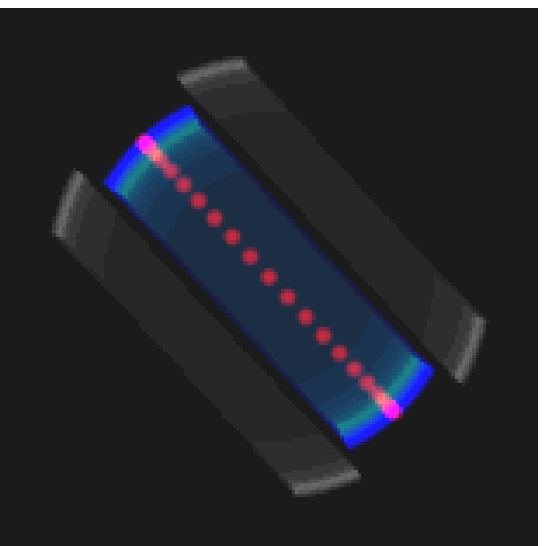}
\hspace{0.15in}
\includegraphics[angle=0,scale=1.00]{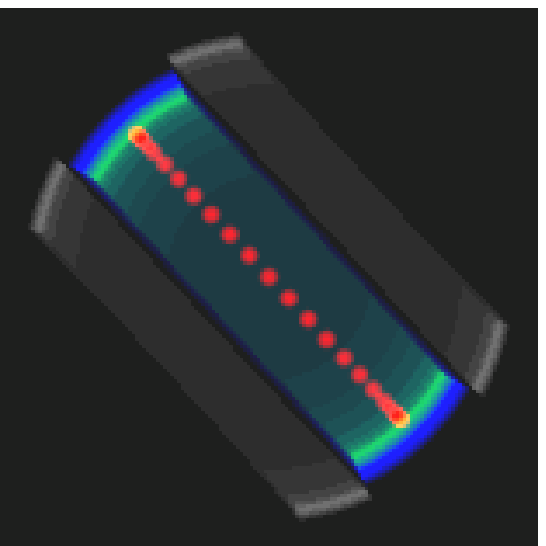} \\
\vspace{0.05in}
\includegraphics[angle=0,scale=1.0]{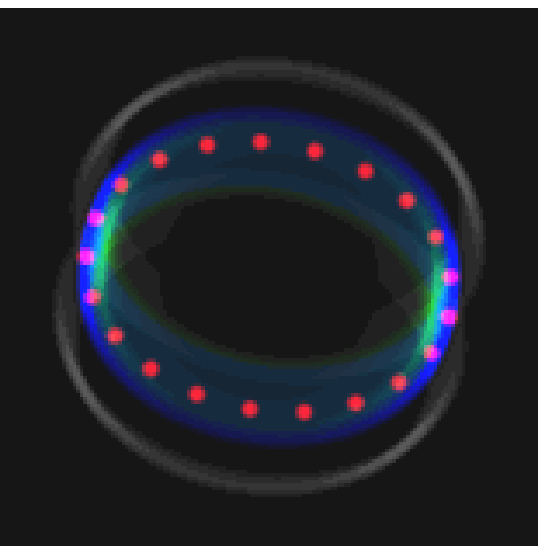}
\hspace{0.15in}
\includegraphics[angle=0,scale=1.00]{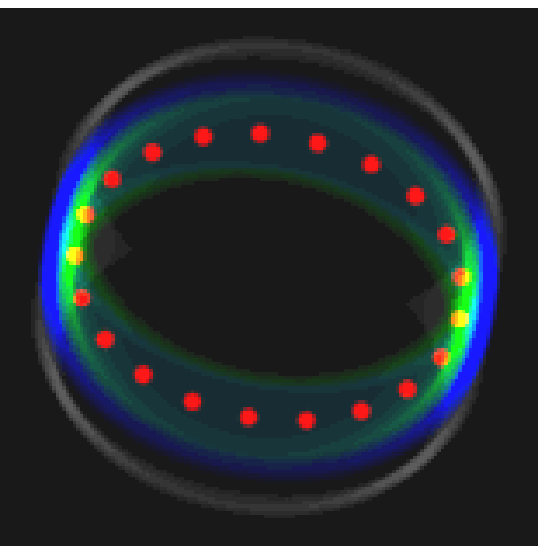} \\
\end{center}
\caption{Hydrodynamics-based geometric emission models of \sna\ at epochs
2004.5 (left column) and 2011.5 (right column).  Three views are shown: a ``polar''
view (top; looking perpendicular to the equatorial plane),
a ``side'' view (middle; north is up and the Earth is to the left),
and the usual sky view (bottom; north up, east to left).
Four emission components are color-coded: the shocked \hii\ material (blue),
the reverse-shocked ejecta (green), the emission from the shocked
ER (red spots), and higher-lattitude 8.6~GHz radio emission (gray).
\label{fig:hydrogeom}}
\end{figure*}

We can construct schematic geometric visualizations of \sna\ using
our simple hydrodynamic models to set the locations and fluxes 
of the components at a given epoch.
The main emission components at two different epochs
are shown in multiple views in Figure~\ref{fig:hydrogeom}
using a color-coding similar to that for the shocked plasma in the schematic of
Figure~\ref{fig:rsscompare}.
These images were created using
simple software extensions to ISIS \citep{Dewey09}\,\footnote{\,See
also: {\tt http://space.mit.edu/home/dd/Event2D/}}.

The \hii\ (``no ER'') simulation gives rise to emission from
i) shocked \hii\ material and ii) reverse-shocked ejecta; these
regions extend out of the ER plane as shown in the images.
The ``with ER'' simulation is shown using a schematic geometry of
21 discrete spots to emphasize our ``X-ray hot spots'' conclusion that
is based on the small value of $\Omega_{\rm ER}$.
The ``with ER'' emission components include: iii)
the shocked dense ER and its clumps, iv) the shocked \hii\ material that is
in the plane of the ER, and v) reverse-shocked ejecta also in
the ER plane.
Because the effective solid angle associated with the ER simulation is so small
($\Omega_{\rm ER}/4\pi=$~0.0016) these last two components make only
minor flux contributions which are exceeded by the first
three.  We note that component ``iv)'' is the plasma that
is ``shocked again'' in the reflected shock structure (RSS)
paradigm \citep{Zhekov09,Zhekov10};
our ``with ER'' hydrodynamics does show
the reflected shock(s) having velocity effects on the shocked \hii\ region
interior to the shocked ER for
about 1.5 years, after which the region's velocity recovers and is similar to
other regions, e.g., see the velocity curve in
Figure~\ref{fig:modprofile}, bottom panel.

These visualizations suggest the need for fitting the X-ray
imaging data with multiple spatial components
and provide some guidance on the component properties.
Recent work of \citet{Ng09} has explored a
two-component spatial model in fitting \sna's image, and
as they emphasize, further observations are needed to come
to a firm conclusion.

\clearpage
%%%%%%%%%%%%%%%%%%%%%%%%%%%%%%%%%%%%%%%%%%%%%%%%%%%%%%%%%%%%%%%%%%%%%%%%%%%%%%
\section{CONCLUSIONS}

In this paper we have modelled the X-ray emission from \sna,
focussing on the very-broad (\vb) component seen with the
high-resolution grating instruments. 
Although the CSM of \sna\ is known to have a complex
shape from optical observations, we have chosen to model it using 1D
simulations in and out of the equatorial plane, our \hii\ and ER
hydrodynamics.  Even with these simplifying
assumptions the models yield a good fit to the X-ray-measured radii
and growth rates, the multi-band X-ray light curves, and
the high-resolution X-ray spectra.

Our results show that better fits to the X-ray spectra over the last
decade are obtained if a very-broad component of emission, with a FWHM of
about 9300 km s$^{-1}$, is included. Such a \vb\ component is present
in the data over at least the last decade.
Our hydrodynamic simulations provide
a natural explanation for this component: it arises from the shocked \hii\
material and it is the dominant X-ray component
until $\sim$\,5500~days after the SN explosion.
Since then the total 0.5--2~keV flux dramatically
increased due to the ``ER'' collision, yet
the inferred 0.5--2~keV flux of the \vb\ component itself
continued to grow in agreement with our hydrodynamics expectations.
Identifying the 3--10~keV flux as originating primarilly from the
\vb\ component provides a natural explanation for the difference in the
0.5--2 and 3--10~keV light curves.
At the present epoch, the \vb\ contribution is $\sim$\,20\% of the
total 0.5--2~keV flux, and of this the hydrodynamics suggests that
roughly half is coming from reverse-shocked {\it ejecta}.
This ejecta contribution should grow and may cause
the \vb\ component to become selectively
enhanced in lines that reflect the (outer) ejecta composition.

Our ER hydrodynamics adquately reproduces the observed X-ray image radii
and the bulk of the late-time 0.5--2~keV flux increase.
We do require clumps in the ER with density enhancements of
$5.5\times$ the $\sim$\,10$^4$\amucc\ ambient ER values, not
unlike the clumps that \citet{Zhekov10} have added within their smooth
CSM density profile.
These are approximations to the actual 2D/3D density structure of the
shock-ER interaction; 
it would not be surprising if the effective clumping parameters might change
in time, e.g., as clumps evaporate and/or turbulent structures evolve.

Our approach allows us to make predictions of what the image size, X-ray spectrum
and light curve will look like in the next few years, in particular
for the two cases of i) a continuing on-going interaction with dense ER material
(our ``thick ER'' case) and ii) the drop in flux if the ER is
``thin''.  In this latter case, the FS in the
equatorial plane has generally exited the densest parts of the ER
and has begun moving into lower-density material as suggested by
\citet{Park11}.  For this case our hydrodynamics shows a decrease in
the 0.5--2~keV flux of $\sim$\,17\,\% per year.
Note that a third option that cannot be completely ruled out
is that the FS will encounter {\it larger amounts} 
of dense ER material, e.g., the main body of the ER; this would
result in even greater increases in flux than we have shown here.
Comparing these and more realistic multi-dimensional
predictions with future observations will provide significant
information regarding the thickness, density, and structure of the equatorial ring
and the \hii\ region within and outside the equatorial plane
as \sna\ moves into a new phase.

\acknowledgments

We thank the anonymous referee for useful suggestions and for holding our statistical feet
to the fire regarding the significance of the \vb\ component measurements.
We thank David Burrows for access to the ``hetg-11'' data in advance
of their public release.  R.S.\ acknowledges support from the German
Bundesministerium f\"ur Wirtschaft und Technologie~/~Deutsches Zentrum
f\"ur Luft- und Raumfahrt (BMWI/DLR) grant FKZ 50 OR 0907. Support for
this work was provided by the National Aeronautics and Space
Administration (NASA) through Chandra Award Number TM9-0004X to V.V.D.\
at the University of Chicago issued by the \chandra\ X-ray Center (CXC),
which is operated by the Smithsonian Astrophysical Observatory (SAO) for and
on behalf of NASA under contract NAS8-03060.
NASA also provided support through the SAO contract
SV3-73016 to MIT for support of the CXC and Science Instruments.

{\it Facilities:} \facility{CXO (HETG)}.

%%%%%%%%%%%%%%%%%%%%%%%%%%%%%%%%%%%%%%%%%%%%%%%%%%%%%%%%%%%%%%%%%%%%%%%%%%%%%%
% Appendix
\appendix
\section{Spectral Models and the {\tt gsmooth} Function}
\label{sec:gsmooth}

The 3-shock model used in \S\,\ref{sec:3shock}
makes use of the XSPEC spectral library and is defined in ISIS with the command:
\begin{equation}\label{eq:3shock}
{\tt fit\_fun(~"phabs(1)*gsmooth(1,\,vpshock(1)+vpshock(2)+vpshock(3)\,)"~); }
\end{equation}
\noindent where the trailing ``;'' is the S-Lang end-of-expression
character.  The {\tt gsmooth(1)} function here is
used to add blur to the model in order to account for spatial and spectral
effects that are unique to \sna; the {\tt gsmooth} parameters
{\tt Sig@6keV} and {\tt Index} and their values for this purpose
are described further below.

A similar expression is used to define the \vb\ 3-shock model
used in \S\,\ref{sec:vb} as the sum of the original 3-shock model
and a very-broad version of itself:
\begin{eqnarray}\label{eq:vb}
 & & {\tt fit\_fun(~~"setvb(1)*phabs(1)*"~~+} \\
 & & ~~{\tt "(~(1-fracbroad)*gsmooth(1,vpshock(1)+vpshock(2)+vpshock(3))+"~~+} \nonumber \\
 & & ~~ ~~~~~~~~ {\tt
 "fracbroad*gsmooth(2,vpshock(1)+vpshock(2)+vpshock(3))~)"~~~); } \nonumber
\end{eqnarray}
\noindent  Here 
the S-Lang variable, {\tt fracbroad}, sets the fraction of the total which is in the
\vb\ component and
the second {\tt gsmooth} term sets the width of the \vb\ emission.
For a \vb\ component having a FWHM of $v_\svb$, the {\tt Sig@6keV}
parameter is set to $6v_\svb/(2.35c)$ and
the {\tt Index} parameter is set to 1, as discussed further below.
The model component ({\tt setvb}) is a dummy function which returns
unity in the spectral model but is used to control the broad-fraction variable
from an ISIS fit parameter, {\tt setvb(1).fb}.  This is accomplished
before the model is defined with a few lines of code:

\noindent{\tt public variable fracbroad;}  \\
{\tt 
\noindent \%} Define a dummy model function to set the broad fraction \\
{\tt define setvb\_fit (l,h,p) \\
\{  \\
\indent \%} set fracbroad from the input parameter:  \\
{\tt \indent fracbroad = p[0];  \\
\indent return 1;  \\
\} \\
add\_slang\_function ("setvb", ["fb"]);
} \\

\noindent In the remainder of this section we provide more specifics on the
{\tt gsmooth} function and its parameters.

\input{tab_gsmooth}

The {\tt gsmooth} function convolves a spectral model with a Gaussian function
whose width, $\sigma_E$, varies with energy according to:
\begin{equation}
\sigma_E = \sigma_6~(E_{\rm keV}/6.0)^{\alpha}~.
\end{equation}
\noindent The constants of this equation, $\sigma_6$ and $\alpha$,
are set through the values of the {\tt gsmooth}
parameters {\tt Sig@6keV} and {\tt Index},
respectively.
This versatile function can be used to approximate instrumental
effects in place of an \rmf.  For the case of $\alpha=0$ it produces
a constant-with-energy blur, e.g., simulating a micro-calorimeter-like
response.  The case of $\alpha=0.5$
gives a resolution, $dE_{\rm FWHM}$, that goes as the
square-root-of-energy,
e.g., as in a proportional-counter or a CCD detector.
For $\alpha=1$ the resolving-power, $E/dE_{\rm FWHM}$, is constant, and
for $\alpha=2$ we have the case of a blur that is constant {\it in
wavelength} as is nearly the case for a grating instrument like the HETG.

The smoothing term can also be used to include additional source spectral
properties, for example, the Doppler blur from a Gaussian velocity spread,
$v_{\rm rms}$, is obtained by setting $\alpha=1$ and
$\sigma_6=6\,v_{\rm rms}/c$~; if the FWHM velocity width is
given, then we have $\sigma_6=6\,v_{\rm FWHM}/(2.35c)$.
It is in exactly this way that we use {\tt gsmooth(2)} in
Expression~\ref{eq:vb}
to produce the \vb\ component in our model.

Finally, we also make use of the {\tt gsmooth} function to
account for additional blurs that are present in the narrow
lines of the \sna\ grating spectra
but are not included in the spectral model ({\tt vpshock}) 
nor in the point-source instrumental responses (\rmf s). 
These blurs have been noted and included in previous grating-data
analyses \citep{Zhekov05,Dewey08,Sturm10} and result from
i) bulk and turbulent motion of the X-ray emitting plasma and
ii) spatial-spectral effects which broaden the observed line widths
due to the spatial size of \sna.  In the context of the {\tt gsmooth}
function the former is accounted for with an $\alpha=1$ behavior
and the latter with an $\alpha=2$ dependence.  The combination of both
effects can be approximated using an intermediate value of $\alpha$
determined by measuring the narrow-component line width at two
energies.  For the analyses here this has been done using
regions including the \oviii\ line and the Mg line (1.28 -- 1.55~keV);
the measured values and resulting {\tt gsmooth} parameters are 
given in Table~\ref{tab:gsmooth}.
Note that the RGS has $\alpha=1$ because \sna\ is spatially unresolved by
\xmm.

%%\clearpage
%% %%%%%%%%%%%%% Bibliography %%%%%%%%%%%%%%

% Bibliography
%  with bibtex:
%%\bibliographystyle{apj}
%%\bibliography{ddrefs}{}
%  already created:

\input{dd87a11_bbl}
%% %%%%%%%%%%%%% TABLES %%%%%%%%%%%%%%

%% Tables should be submitted one per page, so put a \clearpage before
%% Two options are available to the author for producing tables:  the
%% deluxetable environment provided by the AASTeX package or the LaTeX
%% table environment.  Use of deluxetable is preferred.

%% %%%%%%%%%%%%% FIGURES %%%%%%%%%%%%%%

%% the file, uncomment the \plotone, \plottwo, and
%% If you need a layout that cannot be achieved with \plotone or
%% \plottwo, you can invoke the graphicx package directly with the
%% \includegraphics command or use \plotfiddle. For more information,
%% please see the tutorial on "Using Electronic Art with AASTeX" in the
%% documentation section at the AASTeX Web site,
%% http://www.journals.uchicago.edu/AAS/AASTeX.

\end{document}

%% file: tab_gratobs.tex
%
% tab_gratobs.tex 
% Part of D. Dewey et al. (2012)
%
% - Epoch values are from Event2D for HETG,LETG data;
% - SN age is then calc from 365.25*(<Epoch> - 1987.148)
%   For RGS the SN age (days) is from Haberl and Epoch is calculated.
% - Expos. is from isis when data are read in.
% - Fluxes are from flux_plot is isis_11-1-1:
%   dat_name="HETG-99"; .source setup_data  .source load_model  
%   freeze([1,[8:16]]);  fit_counts;    .source flux_plot

\begin{deluxetable}{lccccrlrlrl}
\tablewidth{0pt}
\tabletypesize{\scriptsize}
\tablecaption{High-Resolution X-ray Observations of \sna\ to an Age of 25 Years
  \label{tab:gratobs}}
\tablehead{
   &
\colhead{Epoch\tablenotemark{\,b}} &
\colhead{\obs(s)} &
\colhead{Expos.} &
\colhead{SN Age} &
\multicolumn{2}{c}{$F$(0.47--0.79)} &
\multicolumn{2}{c}{$F$(0.79--1.1)} &
\multicolumn{2}{c}{$F$(1.1--2.1)} \\
% line two
\colhead{Data ID\tablenotemark{\,a}} & 
  (year)  &   & (ks) & (days)\tablenotemark{b}  & 
\multicolumn{2}{c}{(cgs)\tablenotemark{\,c}} &
\multicolumn{2}{c}{(cgs)\tablenotemark{\,c}} &
\multicolumn{2}{c}{(cgs)\tablenotemark{\,c}} 
}
\startdata
HETG-99  & 1999.76  &  124, 1387  & 116.2 & 4606  & 
     0.94 &[12\%]  &  0.57 &[9\%]  &  0.85 &[5\%] \\
ACIS-00\tablenotemark{\,d}  & 2000.93  &  1967  & 98.8 & 5036  & 
     0.67 &[2\%]  &  0.71 &[2\%]  &  1.01 &[2\%] \\
RGS-03   & 2003.36  &  0144530101  & 111.9  & 5920   &
     {4.2}  &  & {3.6} &  & {3.4} &  \\
LETG-04  & 2004.66  &  4640, 5362, 4641, 6099, 5363  & 288.8 &  6396  &
     5.4 &[2\%]  &  6.2 &[2\%]  &  5.2 &[1\%] \\
RGS-07   & 2007.05  &  0406840301  & 110.5  & 7268  &
     {10.0} & & {13.8} & & {12.2} & \\
HETG-07  & 2007.23  &   8523, 8537, 7588, 8538, 7589,    & 354.9 &  7335  &
     13.4 &[3\%]  &  15.5 &[1\%]  & 12.6 &[1\%] \\
         &          &  8539, 8542, 8487, 8543, 8544,     &   &    \\
         &          &  8488, 8545, 8546, 7590  &   &    \\
LETG-07  & 2007.69  &  9581, 9582, 9580, 7620, 7621,   &  285.2 &   7503  &
     13.4 &[2\%]  &  18.2 &[1\%]  &  15.4 &[1\%] \\
         &          &       9591, 9592, 9589, 9590  &  &    \\
RGS-08   & 2008.03  &  0506220101 & 113.7  &  7627   &
     {12.7} & & {18.0} & & {14.5} \\
hetg-08  & 2008.50  &  9144  & 42.0 &  7799  &
     \nodata  & &  19.6 &[3\%]  &  18.3 &[2\%] \\
RGS-09   & 2009.08  & 0556350101  & 101.4  &   8012  &
     {15.3} & & {22.2} & & {17.9}  \\
hetg-09  & 2009.30  &   10221, 10852, 10853, 10854,   &  129.7 &  8091 &
     \nodata & &  23.2 &[2\%]  &  21.2 &[1\%] \\
         &          &       10855; 10222, 10926  &  &     \\
RGS-10   & 2009.93  & 0601200101  & 91.6  &  8320  &
     {16.4} & & {25.2} & & {21.7}  \\
hetg-10  & 2010.46  &  12125, 12126, 11090;  & 118.3 &  8515 &
     \nodata  & &  25.0 &[2\%]  & 24.8 &[1\%] \\
         &          &  13131, 11091  \\
RGS-11   & 2010.93  &  0650420101 & 65.7  &   8687  &
     {17.4} & & {29.0} & & {24.9} \\
HETG-11  & 2011.17  &   12145, 13238, 13239, 12146  &  177.7 &  8774  &
     15.8 &[3\%]  &  27.6 &[1\%]  &  27.4 &[1\%] \\
hetg-11  & 2011.47  &  12539; 12540, 14344  &  101.3 & 8883 &
     \nodata & &  25.2 &[2\%]  &  27.1 &[1\%] \\
%
% Future observations 
%%\hline
% "RGS-12" scheduled for 2 Dec 2011, starting at 07:23. ~ 9048 days.
RGS-12   & $\sim$\,'11.92  & 0671080101 &  $\sim$\,68 &
$\sim$\,9048  &
     \nodata & & \nodata  & & \nodata &  \\
hetg-12  & $\sim$\,'12.22  &  13735, 14417  &  $\sim$\,75 &  $\sim$\,9157 &
    \nodata  & & \nodata &  & \nodata  & \\
\enddata
\tablenotetext{a}{The lower-case ``hetg'' data sets indicate ACIS
monitoring observations where a narrow readout is used; this reduces the 
wavelength coverage of the HETG.}
\tablenotetext{b}{Epoch gives the observation-averaged date of the
data set; SN Age is in days-since-explosion given by  365.25*(\,Epoch $-$
1987.148\,).}
\tablenotetext{c}{$F$($E_1$--$E_2$) is the observed flux from
$E_1$\,keV to $E_2$\,keV in units of $10^{-13}$\,erg/s/cm$^2$.
It is model-independent for the Chandra data sets with the statistical
error given in brackets.
The RGS fluxes are based on the 3-shock model fits (\S\,\ref{sec:3shock});
statistical errors are not given since model-fitting errors of 5--10\% dominate.
}
\tablenotetext{d}{This non-grating ACIS observation is included here
for reference because it is the deepest very-early observation.
}
%
%% Text for table notes should follow after the \enddata but before
%% the \end{deluxetable}. Make sure there is at least one \tablenotemark
%% in the table for each \tablenotetext.
%
\end{deluxetable}

%% file: tab_3shock.tex
%
% tab_3shock.tex 
% Part of D. Dewey et al. (2011)
%

% Values are from /Users/Dan/_Science/SN1987A/isis_110101/Pars_3sh/
% Updated on 2/. . . /12.
%   When updating, also update tab_obs1967.tex 3sh values.

%%\begin{deluxetable}{lcccccccccccc}
\begin{deluxetable}{lccccccccccc}
\rotate
\tablewidth{0pt}
\tabletypesize{\scriptsize} % 8pt
\tablecaption{Parameters of the Global 3-Shock Model Fits
\label{tab:3shock}}
\tablehead{
   &
%%\colhead{Age}  &
  \multicolumn{3}{c}{ Normalizations~\tablenotemark{a} } &
\colhead{$kT_{\rm \,lo}$} &
\colhead{$\tau_{\rm \,mid}$} &
\colhead{$kT_{\rm \,hi}$} &
  \multicolumn{5}{c}{ ---~~~Relative abundances~\tablenotemark{b}~~--- }  \\
\colhead{Data ID} & 
%% (d) &
\colhead{$N_{\rm lo}$} &
\colhead{$N_{\rm 1.15}$} &
\colhead{$N_{\rm hi}$} &
%%      (keV) & (s\,cm$^{-3}$) & (keV) &
      (keV) & (10$^{11}{{\rm s}\over{{\rm cm}^{3}}}$) & (keV) &
%%      (keV) & (10$^{11}$s/cm$^{3}$) & (keV) &
\colhead{N} &
\colhead{O} &
\colhead{Ne} &
\colhead{Si} &
\colhead{Fe}
}
\startdata
% This is the output from running isis_120201/gen_3sh_table.sl .
%  Manual additions: 
%       *** HETG-99 and RGS-03:  put [ ]'s around the 0.74 and 0.50  kT_lo's and remove their errora 
%       Add supressed ``0''s as needed, e.g., 4 --> 4.0,  0.4 --> 0.40
HETG-99  & 0.17 $\pm$35\% & [0]    & 0.13 $\pm$12\%  & [0.74] & 2.1 $\pm$342\% & [4.3]   & [0.56]    & 0.12 $\pm$286\%  & 0.13 $\pm$185\%  & 0.13 $\pm$135\%  & 0.10 $\pm$160\% \\
ACIS-00  & 0.18 $\pm$20\% & [0]    & 0.12 $\pm$9\%  & 0.73 $\pm$11\%  & 8.2 $\pm$19\%  & 4.2 $\pm$14\%  & 1.6 $\pm$63\%  & 0.47 $\pm$50\%  & 0.61 $\pm$42\%  & 0.68 $\pm$27\%  & 0.12 $\pm$45\% \\
\hline
RGS-03  & 3.6 $\pm$12\% & [0]    & 0.14 $\pm$5289\%  & [0.50]   & 4.1 $\pm$46\% & [4.3]    & 0.24 $\pm$41\%  & 0.053 $\pm$34\%  & 0.11 $\pm$21\%  & 0.42 $\pm$184\%  & 0.027 $\pm$23\% \\
RGS-07  & 3.0 $\pm$6\%  & 0.35 $\pm$36\%  & 1.1 $\pm$11\%  & 0.51 $\pm$2\%  & 4.8 $\pm$11\% & [4.3]    & 1.1 $\pm$12\%  & 0.15 $\pm$5\%  & 0.47 $\pm$9\%  & 0.61 $\pm$19\%  & 0.28 $\pm$7\% \\
RGS-08  & 3.4 $\pm$3\%  & 0.72 $\pm$15\%  & 1.3 $\pm$2\%  & 0.52 $\pm$2\%  & 4.7 $\pm$1\% & [2.7]    & 1.1 $\pm$4\%  & 0.15 $\pm$4\%  & 0.40 $\pm$7\%  & 0.37 $\pm$21\%  & 0.30 $\pm$5\% \\
RGS-09  & 5.3 $\pm$3\%  & 1.4 $\pm$9\%  & 1.2 $\pm$12\%  & 0.53 $\pm$1\%  & 7.9 $\pm$6\% & [2.7]    & 1.2 $\pm$0\%  & 0.17 $\pm$0\%  & 0.44 $\pm$5\%  & 0.35 $\pm$19\%  & 0.25 $\pm$3\% \\
RGS-10  & 5.1 $\pm$3\%  & 1.6 $\pm$7\%  & 1.5 $\pm$0\%  & 0.54 $\pm$1\%  & 10 $\pm$0\% & [2.7]    & 1.5 $\pm$0\%  & 0.20 $\pm$0\%  & 0.53 $\pm$5\%  & 0.57 $\pm$7\%  & 0.29 $\pm$2\% \\
RGS-11  & 4.7 $\pm$3\%  & 1.3 $\pm$0\%  & 2.1 $\pm$6\%  & 0.57 $\pm$1\%  & 15 $\pm$0\% & [2.7]    & 2.0 $\pm$0\%  & 0.31 $\pm$0\%  & 0.76 $\pm$0\%  & 0.61 $\pm$16\%  & 0.39 $\pm$2\% \\
\hline
LETG-04  & 2.3 $\pm$4\%  & 0.64 $\pm$10\%  & 0.28 $\pm$8\%  & 0.48 $\pm$2\%  & 3.7 $\pm$8\%  & 5.3 $\pm$9\% & [0.56]    & 0.090 $\pm$9\%  & 0.26 $\pm$6\%  & 0.30 $\pm$6\%  & 0.13 $\pm$6\% \\
HETG-07  & 5.2 $\pm$1\%  & 1.7 $\pm$2\%  & 0.64 $\pm$2\%  & 0.47 $\pm$0\%  & 3.4 $\pm$0\%  & 4.3 $\pm$1\% & [0.56]    & 0.096 $\pm$0\%  & 0.22 $\pm$0\%  & 0.31 $\pm$0\%  & 0.16 $\pm$0\% \\
LETG-07  & 5.2 $\pm$2\%  & 1.4 $\pm$2\%  & 1.2 $\pm$2\%  & 0.48 $\pm$1\%  & 3.7 $\pm$4\%  & 2.4 $\pm$1\% & [0.56]    & 0.090 $\pm$3\%  & 0.27 $\pm$2\%  & 0.34 $\pm$1\%  & 0.19 $\pm$2\% \\
hetg-08  & 6.6 $\pm$7\%  & 1.6 $\pm$37\%  & 1.5 $\pm$31\% & [0.52] & 3.6 $\pm$18\%  & 2.4 $\pm$18\% & [0.56]   & [0.11]    & 0.20 $\pm$10\%  & 0.32 $\pm$8\%  & 0.14 $\pm$10\% \\
hetg-09  & 5.5 $\pm$6\%  & 2.2 $\pm$10\%  & 1.6 $\pm$9\% & [0.52]    & 5.9 $\pm$12\%  & 2.7 $\pm$6\% & [0.56]   & [0.11]    & 0.30 $\pm$7\%  & 0.37 $\pm$5\%  & 0.22 $\pm$5\% \\
hetg-10  & 5.9 $\pm$7\%  & 2.8 $\pm$14\%  & 1.9 $\pm$16\% & [0.52]    & 5.7 $\pm$13\%  & 2.7 $\pm$12\% & [0.56]   & [0.11]    & 0.27 $\pm$9\%  & 0.37 $\pm$5\%  & 0.19 $\pm$5\% \\
HETG-11  & 6.2 $\pm$1\%  & 3.1 $\pm$0\%  & 2.0 $\pm$0\%  & 0.54 $\pm$1\%  & 7.0 $\pm$0\%  & 2.7 $\pm$0\% & [0.56]    & 0.12 $\pm$0\%  & 0.30 $\pm$0\%  & 0.39 $\pm$0\%  & 0.20 $\pm$0\% \\
hetg-11  & 4.9 $\pm$8\%  & 3.5 $\pm$14\%  & 1.9 $\pm$21\% & [0.54] & 7.9 $\pm$18\%  & 3.1 $\pm$19\% & [0.56]   & [0.12]    & 0.32 $\pm$9\%  & 0.40 $\pm$5\%  & 0.20 $\pm$6\% \\
\enddata
\tablecomments{Parameter values are followed by an indication of their $\pm$\,1-$\sigma$ statistical
confidence range, $c_{\rm lo}$ to $c_{\rm hi}$, given as a percentage:
$ 100(\sqrt{c_{\rm hi}/c_{\rm lo}}-1$).
A bracketed parameter value indicates that
it has been fixed at that value during fitting; these fixed values are
generally based on other datasets' results.
$N_{\rm 1.15}$ is not used (set to 0) for the three earliest data sets,
due to their lower total source counts.
$kT_{\rm \,lo}$ is fixed for the HETG-99, RGS-03,
and the ``hetg'' datasets which have reduced low-energy counts or coverage.
$kT_{\rm \,hi}$ is fixed for the RGS and HETG-99 datasets
because of reduced high-energy coverage and counts.
The N abundance is only fit for the RGS and ACIS-00 datasets,
and the O abundance is fixed for the
``hetg'' datasets which have reduced low-energy coverage and counts.
}
% Norms:  Xnorm = 1.e-14 * \int nenHdV / (4\piD^2) 
%                        cm^-3 cm^-3 cm^3 cm^-2 = cm^-5
\tablenotetext{a}{The $N_x$ values given in the table are 10$^3$ times
the XSPEC {\tt vpshock} models' ``norm'' parameters, with units of
cm$^{-5}$.  Specifically, we have:
$N_x$ = 1\tttt{-11}~$ (\int n_e n_H dV) / (4\pi D^2) $
with number densities ($n_e$, $n_H$), volume ($dV$),
and source distance ($D$) in cgs units.
}
\tablenotetext{b}{Abundances are relative to the solar photospheric
values of \citet{AG89}.  Abundances of Mg and S (not tabulated) are
similar to those of Ne and Si, respectively.
$N_H$ has been fixed at 1.3\tttt{21}~cm$^{-2}$. }
\end{deluxetable}

%% file: tab_vb9300.tex
%
% tab_vb9300.tex 
% Part of D. Dewey et al. (2012)
%

% Values are generated by `/isis_120201/gen_vbstats_table.sl
% getting data from .txt files there (whose values come from
% results in `/vb_120201/dia_vb.sl).
%
% Updated 3/12/12.

\begin{deluxetable}{lcrrrrrrc}
\tablewidth{0pt}
\tabletypesize{\scriptsize}
\tablecaption{Very-Broad Fitting Statistics and Results
\label{tab:vbstats}}
\tablehead{
     &
\colhead{$N_{\rm data}$}  &
  &
%%\multicolumn{2}{c}{---~~\vb\ Fit~~---} &
  &  &
\multicolumn{4}{c}{---~~$v_\svb\,=$~9300\kms\ FWHM~~---}  \\
\colhead{Data ID} & 
  $-N_{\rm free}$  &
{$\chi^2_\svb$}   &
{$\Delta\chi^2_2$}   &
{$F_2$} &
{$\Delta\chi^2_1$}   &
{$F_1$} &
\multicolumn{2}{c}{$f_\svb$~~\&~~$1$-$\sigma$ range}
}
\startdata
%  Changes to the s/w output:
%  - change HETG-99 to:  HETG-99\tablenotemark{\,a}
%  - add trailing .0's to third -- eighth columns
%  - add trailing 0s to last 3 values.
HETG-99\tablenotemark{\,a} & 32$-$9  & 24.0  & 2.0  & 1.0  & 1.2  & 1.1  & 0.372  & 0.02 -- 0.63  \\
RGS-03 & 226$-$8  & 352.7  & 86.0  & 26.6  & 80.0  & 48.9  & 0.655  &
0.61 -- 0.70  \\
LETG-04 & 130$-$9  & 174.3  & 68.6  & 23.8  & 66.9  & 46.4  & 0.248  & 0.22 -- 0.28  \\
RGS-07 & 216$-$9  & 360.5  & 36.3  & 10.4  & 31.0  & 17.7  & 0.175  &
0.14 -- 0.20  \\
HETG-07 & 622$-$10  & 1146.1  & 347.8  & 92.9  &  \nodata   &  \nodata   & 0.244  & 0.23 -- 0.26  \\
LETG-07 & 140$-$9  & 355.2  & 81.7  & 15.1  & 48.6  & 16.5  & 0.150  & 0.13 -- 0.17  \\
RGS-08 & 231$-$9  & 450.3  & 71.6  & 17.6  & 45.7  & 21.4  & 0.181  & 0.16 -- 0.21  \\
hetg-08 & 146$-$9  & 174.4  & 28.9  & 11.4  & 26.4  & 20.6  & 0.254  &
0.21 -- 0.30  \\
RGS-09 & 217$-$9  & 416.1  & 68.9  & 17.2  & 33.7  & 15.6  & 0.158  & 0.13 -- 0.18  \\
hetg-09 & 349$-$9  & 546.6  & 76.3  & 23.7  & 75.0  & 46.7  & 0.208  & 0.18 -- 0.23  \\
RGS-10 & 222$-$9  & 404.0  & 81.2  & 21.4  & 66.8  & 34.2  & 0.206  & 0.18 -- 0.23  \\
hetg-10 & 328$-$9  & 427.0  & 44.5  & 16.6  & 42.4  & 31.6  & 0.178  &
0.15 -- 0.20  \\
RGS-11 & 201$-$9  & 429.6  & 67.1  & 15.0  & 57.5  & 25.2  & 0.211  & 0.19 -- 0.24  \\
HETG-11 & 599$-$10  & 1115.3  & 116.7  & 30.8  &  \nodata   &  \nodata   & 0.173  & 0.16 -- 0.19  \\
hetg-11 & 319$-$9  & 422.0  & 47.2  & 17.3  & 45.1  & 33.1  & 0.201  & 0.17 -- 0.23  \\
\enddata
\tablecomments{The four left-most numerical columns give values related
to the very-broad fitting of the data sets: the number of data bins
minus the number of free fit parameters, the $\chi^2$ value for the
\vb\ fit, the amount by which $\chi^2_\svb$ is an improvement over
(less than) the non-\vb\
$\chi^2$ value, and the $F$-statistic for adding the 2
\vb\ degrees of freedom to the model.  Specifically,
$F_2=(\Delta\chi^2_2/2)/(\chi^2_\svb/(N_{\rm data}-N_{free}))$ and
values of $F_2$ greater than $F_{\rm crit}\approx$~7.3 indicate a high
confidence, $>0.999$, that the addition of the 2  \vb\ parameters does
result in a significantly improved model. \\
~~~~The right-most four columns give values related to a constrained \vb\ fitting where
the \vb\ velocity width is fixed at the HETG-07/HETG-11
average value; thus there is only a single free \vb\ parameter: $f_\svb$, the \vb\ fraction.
The first two of these columns give the $\chi^2$ improvement and the
appropriate $F$-statistic:
$F_1=(\Delta\chi^2_1/1)/(\chi^2_{9300}/(N_{\rm data}-N_{free}+1))$ with
$\chi^2_{9300} = \chi^2_\svb + \Delta\chi^2_2 - \Delta\chi^2_1$.
In this case we have $F_{\rm crit}\approx$~11.4.  The right-most two
columns give the best-fit \vb\ fraction and its 1-$\sigma$ confidence
range for the constrained fitting.
}
\tablenotetext{a}{The HETG-99 values tabulated here are from the same analysis as
used for the other data sets in this table.  The ``stacked analysis''
result of \citet{Michael02}, $f_\svb = 0.78\,\pm\,0.1$, is considered more accurate
and is used for the very-broad HETG-99 point in Fig.~\ref{fig:modlc}}
%% Text for table notes should follow after the \enddata but before
%% the \end{deluxetable}. Make sure there is at least one \tablenotemark
%% in the table for each \tablenotetext.
%
\end{deluxetable}

%% file: tab_3shVB.tex
%
% tab_3shVB.tex 
% Part of D. Dewey et al. (2011)
%

% Values are from /Users/Dan/_Science/SN1987A/isis_120201/Pars_3sh/
% Updated on 2/. . . /12.
%   When updating, also update tab_obs1967.tex 3sh values.

%%\begin{deluxetable}{lcccccccccccc}
\begin{deluxetable}{lccccccccccc}
\rotate
\tablewidth{0pt}
\tabletypesize{\scriptsize} % 8pt
\tablecaption{Parameters of the Global 3-Shock-with-Very-Broad-Component Model Fits
\label{tab:3shVB}}
\tablehead{
   &
%%\colhead{Age}  &
  \multicolumn{3}{c}{ Normalizations~\tablenotemark{a} } &
\colhead{$kT_{\rm \,lo}$} &
\colhead{$\tau_{\rm \,mid}$} &
\colhead{$kT_{\rm \,hi}$} &
  \multicolumn{5}{c}{ ---~~~Relative abundances~\tablenotemark{b}~~--- }  \\
\colhead{Data ID} & 
%% (d) &
\colhead{$N_{\rm lo}$} &
\colhead{$N_{\rm 1.15}$} &
\colhead{$N_{\rm hi}$} &
%%      (keV) & (s\,cm$^{-3}$) & (keV) &
      (keV) & (10$^{11}{{\rm s}\over{{\rm cm}^{3}}}$) & (keV) &
%%      (keV) & (10$^{11}$s/cm$^{3}$) & (keV) &
\colhead{N} &
\colhead{O} &
\colhead{Ne} &
\colhead{Si} &
\colhead{Fe}
}
\startdata
% This is the output from running isis_120201/gen_3sh_table.sl .
%  Manual additions: 
%       *** put [ ] around the 0.74 and remove the error in HETG-99  ***
%       ACIS-00 -->  ACIS-00\,\tablenotemark{c}
%       Add supressed ``0''s as needed, e.g., 4 --> 4.0,  0.4 --> 0.40
HETG-99  & 0.10 $\pm$54\% & [0]    & 0.12 $\pm$9\%  & [0.74]  & 9.2 $\pm$211\% & [4.2]   & [0.56]    & 0.82 $\pm$214\%  & 0.77 $\pm$154\%  & 0.45 $\pm$120\%  & 0.26 $\pm$144\% \\
ACIS-00\,\tablenotemark{c}  & 0.18 $\pm$40\% & [0]    & 0.12 $\pm$9\%  & 0.73 $\pm$13\%  & 8.2 $\pm$69\%  & 4.2 $\pm$17\%  & 1.6 $\pm$142\%  & 0.47 $\pm$25\%  & 0.61 $\pm$73\%  & 0.68 $\pm$43\%  & 0.12 $\pm$24\% \\
\hline
RGS-03  & 1.5 $\pm$21\% & [0]    & 0.17 $\pm$91\%  & 0.57 $\pm$8\%  & 3.0 $\pm$24\% & [4.3]    & 0.86 $\pm$29\%  & 0.16 $\pm$23\%  & 0.28 $\pm$16\%  & 0.68 $\pm$145\%  & 0.11 $\pm$25\% \\
RGS-07  & 2.4 $\pm$7\%  & 0.16 $\pm$64\%  & 1.0 $\pm$10\%  & 0.52 $\pm$2\%  & 5.2 $\pm$8\% & [3]    & 1.6 $\pm$7\%  & 0.21 $\pm$6\%  & 0.64 $\pm$8\%  & 0.81 $\pm$15\%  & 0.36 $\pm$6\% \\
RGS-08  & 2.6 $\pm$2\%  & 0.40 $\pm$21\%  & 1.2 $\pm$6\%  & 0.53 $\pm$2\%  & 5.7 $\pm$1\% & [2.7]    & 1.9 $\pm$3\%  & 0.25 $\pm$4\%  & 0.62 $\pm$7\%  & 0.54 $\pm$20\%  & 0.43 $\pm$4\% \\
RGS-09  & 4.3 $\pm$3\%  & 0.86 $\pm$9\%  & 1.4 $\pm$9\%  & 0.54 $\pm$1\%  & 8.1 $\pm$8\% & [2.7]    & 1.6 $\pm$5\%  & 0.23 $\pm$2\%  & 0.57 $\pm$6\%  & 0.44 $\pm$18\%  & 0.33 $\pm$4\% \\
RGS-10  & 3.8 $\pm$4\%  & 0.98 $\pm$15\%  & 1.6 $\pm$9\%  & 0.55 $\pm$1\%  & 12 $\pm$3\% & [2.7]    & 2.5 $\pm$4\%  & 0.33 $\pm$3\%  & 0.80 $\pm$5\%  & 0.81 $\pm$10\%  & 0.42 $\pm$3\% \\
RGS-11  & 3.4 $\pm$2\%  & 0.68 $\pm$16\%  & 2.1 $\pm$8\%  & 0.57 $\pm$2\%  & 17 $\pm$8\% & [2.7]    & 3.3 $\pm$2\%  & 0.49 $\pm$1\%  & 1.2 $\pm$5\%  & 0.93 $\pm$14\%  & 0.59 $\pm$0\% \\
\hline
LETG-04  & 1.8 $\pm$6\%  & 0.41 $\pm$11\%  & 0.34 $\pm$7\%  & 0.50 $\pm$3\%  & 4.3 $\pm$4\%  & 4.5 $\pm$14\% & [0.56]    & 0.14 $\pm$10\%  & 0.38 $\pm$7\%  & 0.40 $\pm$7\%  & 0.18 $\pm$5\% \\
HETG-07  & 3.9 $\pm$2\%  & 0.60 $\pm$3\%  & 1.1 $\pm$1\%  & 0.50 $\pm$1\%  & 3.4 $\pm$2\%  & 3.0 $\pm$1\% & [0.56]    & 0.15 $\pm$4\%  & 0.33 $\pm$2\%  & 0.41 $\pm$2\%  & 0.22 $\pm$2\% \\
LETG-07  & 4.6 $\pm$2\%  & 0.84 $\pm$3\%  & 1.5 $\pm$4\%  & 0.49 $\pm$2\%  & 3.8 $\pm$2\%  & 2.3 $\pm$3\% & [0.56]    & 0.11 $\pm$3\%  & 0.33 $\pm$2\%  & 0.39 $\pm$3\%  & 0.22 $\pm$2\% \\
hetg-08  & 5.6 $\pm$9\%  & 0.54 $\pm$189\%  & 1.9 $\pm$25\% & [0.52]    & 4.1 $\pm$15\%  & 2.2 $\pm$15\% & [0.56]   & [0.15]    & 0.32 $\pm$9\%  & 0.42 $\pm$6\%  & 0.19 $\pm$11\% \\
hetg-09  & 4.5 $\pm$6\%  & 1.2 $\pm$23\%  & 1.9 $\pm$10\% & [0.52]    & 5.5 $\pm$13\%  & 2.6 $\pm$6\% & [0.56]   & [0.15]    & 0.41 $\pm$8\%  & 0.46 $\pm$6\%  & 0.28 $\pm$6\% \\
hetg-10  & 5.2 $\pm$7\%  & 1.9 $\pm$22\%  & 2.2 $\pm$15\% & [0.52]    & 5.7 $\pm$17\%  & 2.6 $\pm$9\% & [0.56]   & [0.15]    & 0.35 $\pm$9\%  & 0.46 $\pm$6\%  & 0.23 $\pm$7\% \\
HETG-11  & 5.3 $\pm$2\%  & 2.2 $\pm$1\%  & 2.2 $\pm$1\%  & 0.55 $\pm$1\%  & 7.2 $\pm$2\%  & 2.8 $\pm$2\% & [0.56]    & 0.15 $\pm$7\%  & 0.40 $\pm$2\%  & 0.49 $\pm$2\%  & 0.26 $\pm$2\% \\
hetg-11  & 4.1 $\pm$9\%  & 2.4 $\pm$19\%  & 2.3 $\pm$15\% & [0.55]    & 7.7 $\pm$29\%  & 2.8 $\pm$11\% & [0.56]   & [0.15]    & 0.45 $\pm$14\%  & 0.51 $\pm$9\%  & 0.26 $\pm$7\% \\
\enddata
\tablecomments{The data sets have been fit as before,
Table~\ref{tab:3shock}, with the inclusion of a \vb\ component fixed
to have FWHM = 9300\kms\ and a fraction as given in
Table~\ref{tab:vbstats}.\\
~~~~~~~ Parameter values are followed by an indication of their $\pm$\,1-$\sigma$ statistical
confidence range, $c_{\rm lo}$ to $c_{\rm hi}$, given as a percentage:
$ 100(\sqrt{c_{\rm hi}/c_{\rm lo}}-1$).
A bracketed parameter value indicates that
it has been fixed at that value during fitting; these fixed values are
generally based on other datasets' results.
$N_{\rm 1.15}$ is not used (set to 0) for the three earliest data sets,
due to their lower total source counts.
$kT_{\rm \,lo}$ is fixed for the HETG-99
and the ``hetg'' datasets which have reduced low-energy counts or coverage.
$kT_{\rm \,hi}$ is fixed for the RGS and HETG-99 datasets
because of reduced high-energy coverage and counts.
The N abundance is only fit for the RGS and ACIS-00 datasets,
and the O abundance is fixed for the
``hetg'' datasets which have reduced low-energy coverage and counts.
}
% Norms:  Xnorm = 1.e-14 * \int nenHdV / (4\piD^2) 
%                        cm^-3 cm^-3 cm^3 cm^-2 = cm^-5
\tablenotetext{a}{The $N_x$ values given in the table are 10$^3$ times
the XSPEC {\tt vpshock} models' ``norm'' parameters, with units of
cm$^{-5}$.  Specifically, we have:
$N_x$ = 1\tttt{-11}~$ (\int n_e n_H dV) / (4\pi D^2) $
with number densities ($n_e$, $n_H$), volume ($dV$),
and source distance ($D$) in cgs units.
}
\tablenotetext{b}{Abundances are relative to the solar photospheric
values of \citet{AG89}.  Abundances of Mg and S (not tabulated) are
similar to those of Ne and Si, respectively.
$N_H$ has been fixed at 1.3\tttt{21}~cm$^{-2}$. }
\tablenotetext{c}{A \vb\ fraction of 0.7 was used in fitting
the ACIS-00 data here; the fit results are very similar to
those for ACIS-00 in Table~\ref{tab:3shock} (where \vb\ fraction = 0).
}
\end{deluxetable}

%% file: tab_obs1967.tex
%
% tab_obs1967.tex 
% Part of D. Dewey et al. (2011)
%

% Values are from
% /Users/Dan/_Science/SN1987A/isis_120201/obs_1967_compare.sl
% and referenced par files.

\begin{deluxetable}{cccccccc}
\tablewidth{0pt}
\tabletypesize{\scriptsize}
\tablecaption{Degeneracy of \hii\ Abundances when Fitting ACIS-00
\label{tab:obs1967}}
\tablehead{
  &
\colhead{$kT$(s)} &
\colhead{$\tau$(s)} &
\multicolumn{5}{c}{ ---~~~Relative abundances~\tablenotemark{a} ~~
(1-$\sigma$ range)~~~--- }  \\
\colhead{Model} &
(keV) & ($10^{11}$~s\,cm$^{-3}$) & 
\colhead{N} &
\colhead{O} &
\colhead{Ne} &
\colhead{Si} &
\colhead{Fe}
}
\startdata
% One shock model:
1-shock & 
   2.94--3.34 & 0.94--1.32 &  
   0.49--0.94 & 0.14--0.23 & 0.00--0.17 & 0.28--0.41 &
   0.13--0.25 \\
% ~ Zhekov 2010 2 shock model:
2-shock\,\tablenotemark{b} &
   0.22, 2.8  &  130, 1.4 &  
   0.34--0.72 & 0.11--0.17 & 0.27--0.35 & 0.29--0.38 &
   0.083--0.12 \\
% ``3-shock''
Table~\ref{tab:3shock} &
   0.73, 4.2  & 11.6, 5.8 &  
%  From doing:  more isis_120201/Pars_3sh/Obs_1967_3sh115_AbsconfAbs.dat
   0.89--2.35 & 0.28--0.63 & 0.46--0.92 & 0.59--0.94 &
   0.087--0.18 \\
Hydro-noER &
\multicolumn{2}{c}{14 pairs in Figure~\ref{fig:TTauHII}} & 
   0.51--0.66 & 0.137--0.155 & 0.35--0.42 & 0.32--0.43 &
   0.061--0.080 \\
\enddata
\tablenotetext{a}{Abundances are relative to the solar photospheric
values of \citet{AG89}.    Abundances of Mg and S (not tabulated) are
similar to those of Ne and Si, respectively.  
$N_H$ has been fixed at 1.3\tttt{21}~cm$^{-2}$, unless stated otherwise. }
\tablenotetext{b}{The $kT$ and $\tau$ values for this 2-shock model were estimated from
Figures~3\,\&\,4 of \citet{Zhekov10}, which also gives an $N_H$ of
2.23\tttt{21}~cm$^{-2}$ that was used in the fitting here.}
\end{deluxetable}

%% file: tab_abundsdisc.tex
\begin{deluxetable}{rccccc}
\tablewidth{0pt}
\tabletypesize{\scriptsize}
\tablecaption{Hydrodynamics-based Abundances and Others
\label{tab:abundsdisc}}
\tablehead{
    &
---~~H\,II~~---  &
\colhead{---~~ER\tablenotemark{\,d}~~---} &
\multicolumn{3}{c}{---------~~Other Abundances~~---------} \\
\colhead{Parameters} &
\colhead{ACIS-00} &
\colhead{HETG-\,07\,\&\,11} &
\colhead{Zhekov(`09)\tablenotemark{\,a}} &
\colhead{Zhekov(`10)\tablenotemark{\,b}} &
\colhead{Mattila(`10)\tablenotemark{\,c}} 
}
\startdata
%         ACIS           Nom           
%   Zhekov'09        Zhekov'10             Mattila
N &   0.584\,$\pm${0.07} &   3.3\,$\pm$1.2    &
    0.56\pms{0.09}{0.06}    & 0.32\,$\pm${0.04}    &    2.5\,$\pm$1.0  \\
O &   0.144\pms{0.011}{0.007} &  0.382\,$\pm$0.08   &
    0.081\,$\pm${0.01} & 0.07\,$\pm${0.01}    &   0.22\,$\pm$0.04     \\
Ne &  0.381\,$\pm${0.03} &  0.566\,$\pm$0.03   &
    0.29\,$\pm${0.02}    & 0.22\,$\pm${0.01}    &   0.69     \\
Mg &  0.270\,$\pm${0.04} &  0.394\,$\pm$0.02   &
    0.28\pms{0.01}{0.02}    & 0.17\,$\pm${0.01}    &  \nodata   \\
Si &  0.373\,$\pm${0.05} &  0.469\,$\pm$0.02   &
    0.33\pms{0.02}{0.01}    & 0.26\,$\pm${0.01}    &  \nodata   \\
S &   0.293\pms{0.12}{0.10} &  0.556\,$\pm$0.08   &
    0.30\,$\pm${0.06}    & 0.36\,$\pm${0.02}    &   0.81\,$\pm$0.4     \\
Ar &   0.00\pms{0.23}{0.0}  &  1.2\,$\pm$0.5   &
     [0.537]    & [0.537]    &    0.47    \\
Fe &  0.070\pms{0.010}{0.009} &  0.225\,$\pm$0.02   &
    0.19\pms{0.02}{0.00}    & 0.10\,$\pm${0.01}    &   0.20\,$\pm$0.1  \\
$f_{clump}$     & \nodata & 0.30\,$\pm$0.02   &
    \nodata & \nodata & \nodata \\
$\eta_{clump}$  & \nodata & 5.51\,$\pm$0.45   &
    \nodata & \nodata  & \nodata \\
\enddata
\tablecomments{Abundances are relative to the solar photospheric
values of \citet{AG89}, AG89.  Values in brackets were fixed; those
not tabulated were set to: H=1, He=2.57,
C=0.03, Ar=0.03, Ca=0.03, and Ni=0.07.  $N_H$ has been fixed at
1.3\tttt{21}~cm$^{-2}$. }
\tablenotetext{a}{Values are from \citet{Zhekov09} Table~1 for fits to LETG and
HETG data.}
\tablenotetext{b}{Values are from \citet{Zhekov10} Table~1 for fits to 14 ACIS
spectra; a value of $N_H$=2.23\,$\pm$0.05 \tttt{21}~cm$^{-2}$ was
determined and used.}
\tablenotetext{c}{Values are from \citet{Mattila10}, Table~7,
converted to AG89 equivalents.  Other of their values are:
He=1.74\,$\pm${0.61}, C
undefined, and Ca=1.41.}
\tablenotetext{d}{The error ranges for the ER abundances and clumping
parameters include the difference between the HETG-07 and HETG-11 best-fit values and their
1-$\sigma$ ranges.  The N abundance was set based on
fitting the RGS-11 data over the 1-$\sigma$ range for $\eta_{\rm clump}$.}
\end{deluxetable}

%% file: tab_gsmooth.tex
% file: tab_gsmooth.tex
% Part of D. Dewey et al. (2011)

\begin{deluxetable}{ccccc}
\tablewidth{0pt}
\tabletypesize{\scriptsize}
\tablecaption{Spatial-Spectral Blur Parameters
\label{tab:gsmooth}}
\tablehead{
   &
\colhead{$\sigma_{\rm O\,VIII}$} &
\colhead{$\sigma_{\rm Mg}$} &
    &
\colhead{$\sigma_6$, {\tt Sig@6keV}}  \\
\colhead{Grating} &
(eV) &
(eV) &
\colhead{$\alpha$, {\tt Index}} &
(keV)
}
\startdata
 RGS & 0.628  & [1.36] & 1.0000 & 0.00580 \\
HETG & 0.881  &  2.14  & 1.1454 & 0.01124 \\
LETG & 1.893  &  5.90  & 1.4682 & 0.04946 \\
\enddata
% O VIII is 0.65 keV, ``Mg'' is 1.41 keV.
\end{deluxetable}

%% file: dd87a11.bbl
\begin{thebibliography}{55}
\expandafter\ifx\csname natexlab\endcsname\relax\def\natexlab#1{#1}\fi

\bibitem[{{Anders} \& {Grevesse}(1989)}]{AG89}
{Anders}, E., \& {Grevesse}, N. 1989, \gca, 53, 197

\bibitem[{{Borkowski} {et~al.}(1997{\natexlab{a}}){Borkowski}, {Blondin}, \&
  {McCray}}]{BBMcC97let}
{Borkowski}, K.~J., {Blondin}, J.~M., \& {McCray}, R. 1997{\natexlab{a}},
  \apjl, 476, L31

\bibitem[{{Borkowski} {et~al.}(1997{\natexlab{b}}){Borkowski}, {Blondin}, \&
  {McCray}}]{BBMcC97}
---. 1997{\natexlab{b}}, \apj, 477, 281

\bibitem[{{Borkowski} {et~al.}(2001){Borkowski}, {Lyerly}, \&
  {Reynolds}}]{Borkowski01}
{Borkowski}, K.~J., {Lyerly}, W.~J., \& {Reynolds}, S.~P. 2001, \apj, 548, 820

\bibitem[{{Canizares} {et~al.}(2000){Canizares}, {Huenemoerder}, {Davis},
  {Dewey}, {Flanagan}, {Houck}, {Markert}, {Marshall}, {Schattenburg},
  {Schulz}, {Wise}, {Drake}, \& {Brickhouse}}]{Canizares00}
{Canizares}, C.~R., {et~al.} 2000, \apjl, 539, L41

\bibitem[{{Canizares} {et~al.}(2005){Canizares}, {Davis}, {Dewey}, {Flanagan},
  {Galton}, {Huenemoerder}, {Ishibashi}, {Markert}, {Marshall}, {McGuirk},
  {Schattenburg}, {Schulz}, {Smith}, \& {Wise}}]{Canizares05}
---. 2005, \pasp, 117, 1144

\bibitem[{{Chevalier}(1982)}]{Chevalier82selfsim}
{Chevalier}, R.~A. 1982, \apj, 258, 790

\bibitem[{{Chevalier}(1992)}]{Chevalier92}
---. 1992, \nat, 355, 617

\bibitem[{{Chevalier} \& {Dwarkadas}(1995)}]{Chevalier95}
{Chevalier}, R.~A., \& {Dwarkadas}, V.~V. 1995, \apjl, 452, L45

\bibitem[{{Chevalier} \& {Liang}(1989)}]{Chevalier89}
{Chevalier}, R.~A., \& {Liang}, E.~P. 1989, \apj, 344, 332

\bibitem[{{Colella} \& {Woodward}(1984)}]{Colella84}
{Colella}, P., \& {Woodward}, P.~R. 1984, Journal of Computational Physics, 54,
  174

\bibitem[{{den Herder} {et~al.}(2001){den Herder}, {Brinkman}, {Kahn},
  {Branduardi-Raymont}, {Thomsen}, {Aarts}, {Audard}, {Bixler}, {den Boggende},
  {Cottam}, {Decker}, {Dubbeldam}, {Erd}, {Goulooze}, {G{\"u}del}, {Guttridge},
  {Hailey}, {Janabi}, {Kaastra}, {de Korte}, {van Leeuwen}, {Mauche},
  {McCalden}, {Mewe}, {Naber}, {Paerels}, {Peterson}, {Rasmussen}, {Rees},
  {Sakelliou}, {Sako}, {Spodek}, {Stern}, {Tamura}, {Tandy}, {de Vries},
  {Welch}, \& {Zehnder}}]{denHerder01}
{den Herder}, J.~W., {et~al.} 2001, \aap, 365, L7

\bibitem[{{Dewey} \& {Noble}(2009)}]{Dewey09}
{Dewey}, D., \& {Noble}, M.~S. 2009, in Astronomical Society of the Pacific
  Conference Series, Vol. 411, Astronomical Data Analysis Software and Systems
  XVIII, ed. {D.~A.~Bohlender, D.~Durand, \& P.~Dowler}, 234--238

\bibitem[{{Dewey} {et~al.}(2008){Dewey}, {Zhekov}, {McCray}, \&
  {Canizares}}]{Dewey08}
{Dewey}, D., {Zhekov}, S.~A., {McCray}, R., \& {Canizares}, C.~R. 2008, \apjl,
  676, L131

\bibitem[{{Dwarkadas}(2007{\natexlab{a}})}]{Dwarkadas07RevMex}
{Dwarkadas}, V.~V. 2007{\natexlab{a}}, in Revista Mexicana de Astronomia y
  Astrofisica Conference Series, Vol.~30, Revista Mexicana de Astronomia y
  Astrofisica Conference Series, 49--56

\bibitem[{{Dwarkadas}(2007{\natexlab{b}})}]{Dwarkadas07AIP}
{Dwarkadas}, V.~V. 2007{\natexlab{b}}, in American Institute of Physics
  Conference Series, Vol. 937, Supernova 1987A: 20 Years After: Supernovae and
  Gamma-Ray Bursters, ed. {S.~Immler, K.~Weiler, \& R.~McCray}, 120--124

\bibitem[{{Dwarkadas} {et~al.}(2010){Dwarkadas}, {Dewey}, \&
  {Bauer}}]{Dwarkadas10}
{Dwarkadas}, V.~V., {Dewey}, D., \& {Bauer}, F. 2010, \mnras, 407, 812

\bibitem[{{Dwarkadas} \& {Gruszko}(2012)}]{Dwarkadas12}
{Dwarkadas}, V.~V., \& {Gruszko}, J. 2012, \mnras, 419, 1515

\bibitem[{{Gaensler} {et~al.}(1997){Gaensler}, {Manchester}, {Staveley-Smith},
  {Tzioumis}, {Reynolds}, \& {Kesteven}}]{Gaensler97}
{Gaensler}, B.~M., {Manchester}, R.~N., {Staveley-Smith}, L., {Tzioumis},
  A.~K., {Reynolds}, J.~E., \& {Kesteven}, M.~J. 1997, \apj, 479, 845

\bibitem[{{Ghavamian} {et~al.}(2007){Ghavamian}, {Laming}, \&
  {Rakowski}}]{Ghavamian07}
{Ghavamian}, P., {Laming}, J.~M., \& {Rakowski}, C.~E. 2007, \apjl, 654, L69

\bibitem[{{Haberl} {et~al.}(2006){Haberl}, {Geppert}, {Aschenbach}, \&
  {Hasinger}}]{Haberl06}
{Haberl}, F., {Geppert}, U., {Aschenbach}, B., \& {Hasinger}, G. 2006, \aap,
  460, 811

\bibitem[{{Hasinger} {et~al.}(1996){Hasinger}, {Aschenbach}, \&
  {Truemper}}]{Hasinger96}
{Hasinger}, G., {Aschenbach}, B., \& {Truemper}, J. 1996, \aap, 312, L9

\bibitem[{{Houck} \& {Denicola}(2000)}]{Houck00}
{Houck}, J.~C., \& {Denicola}, L.~A. 2000, in Astronomical Society of the
  Pacific Conference Series, Vol. 216, Astronomical Data Analysis Software and
  Systems IX, ed. {N.~Manset, C.~Veillet, \& D.~Crabtree}, 591

\bibitem[{{Huenemoerder} {et~al.}(2011){Huenemoerder}, {Mitschang}, {Dewey},
  {Nowak}, {Schulz}, {Nichols}, {Davis}, {Houck}, {Marshall}, {Noble},
  {Morgan}, \& {Canizares}}]{TGCat11}
{Huenemoerder}, D.~P., {et~al.} 2011, \aj, 141, 129

\bibitem[{{Inoue} {et~al.}(1991){Inoue}, {Hayashida}, {Itoh}, {Kondo},
  {Mitsuda}, {Takeshima}, {Yoshida}, \& {Tanaka}}]{Inoue91}
{Inoue}, H., {Hayashida}, K., {Itoh}, M., {Kondo}, H., {Mitsuda}, K.,
  {Takeshima}, T., {Yoshida}, K., \& {Tanaka}, Y. 1991, \pasj, 43, 213

\bibitem[{{Itoh} {et~al.}(1987){Itoh}, {Hayakawa}, {Masai}, \&
  {Nomoto}}]{Itoh87}
{Itoh}, H., {Hayakawa}, S., {Masai}, K., \& {Nomoto}, K. 1987, \pasj, 39, 529

\bibitem[{{Larsson} {et~al.}(2011){Larsson}, {Fransson}, {{\"O}stlin},
  {Gr{\"o}ningsson}, {Jerkstrand}, {Kozma}, {Sollerman}, {Challis}, {Kirshner},
  {Chevalier}, {Heng}, {McCray}, {Suntzeff}, {Bouchet}, {Crotts}, {Danziger},
  {Dwek}, {France}, {Garnavich}, {Lawrence}, {Leibundgut}, {Lundqvist},
  {Panagia}, {Pun}, {Smith}, {Sonneborn}, {Wang}, \& {Wheeler}}]{Larsson11}
{Larsson}, J., {et~al.} 2011, \nat, 474, 484

\bibitem[{{Lundqvist}(1999)}]{Lundqvist99}
{Lundqvist}, P. 1999, \apj, 511, 389

\bibitem[{{Luo} \& {McCray}(1991{\natexlab{a}})}]{Luo91a}
{Luo}, D., \& {McCray}, R. 1991{\natexlab{a}}, \apj, 372, 194

\bibitem[{{Luo} \& {McCray}(1991{\natexlab{b}})}]{Luo91b}
---. 1991{\natexlab{b}}, \apj, 379, 659

\bibitem[{{Luo} {et~al.}(1994){Luo}, {McCray}, \& {Slavin}}]{Luo94}
{Luo}, D., {McCray}, R., \& {Slavin}, J. 1994, \apj, 430, 264

\bibitem[{{Masai} {et~al.}(1991){Masai}, {Itoh}, \& {Nomoto}}]{Masai91}
{Masai}, K., {Itoh}, H., \& {Nomoto}, K. 1991, in European Southern Observatory
  Conference and Workshop Proceedings, Vol.~37, European Southern Observatory
  Conference and Workshop Proceedings, ed. {I.~J.~Danziger \& K.~Kjaer}, 197

\bibitem[{{Masai} \& {Nomoto}(1994)}]{Masai94}
{Masai}, K., \& {Nomoto}, K. 1994, \apj, 424, 924

\bibitem[{{Mattila} {et~al.}(2010){Mattila}, {Lundqvist}, {Gr\"{o}ningsson},
  {Meikle}, {Stathakis}, {Fransson}, \& {Cannon}}]{Mattila10}
{Mattila}, S., {Lundqvist}, P., {Gr\"{o}ningsson}, P., {Meikle}, P.,
  {Stathakis}, R., {Fransson}, C., \& {Cannon}, R. 2010, \apj, 717, 1140

\bibitem[{{McCray}(2007)}]{McCray07}
{McCray}, R. 2007, in American Institute of Physics Conference Series, Vol.
  937, Supernova 1987A: 20 Years After: Supernovae and Gamma-Ray Bursters, ed.
  {S.~Immler, K.~Weiler, \& R.~McCray}, 3--14

\bibitem[{{Michael} {et~al.}(2002){Michael}, {Zhekov}, {McCray}, {Hwang},
  {Burrows}, {Park}, {Garmire}, {Holt}, \& {Hasinger}}]{Michael02}
{Michael}, E., {et~al.} 2002, \apj, 574, 166

\bibitem[{{Michael} {et~al.}(2003){Michael}, {McCray}, {Chevalier},
  {Filippenko}, {Lundqvist}, {Challis}, {Sugerman}, {Lawrence}, {Pun},
  {Garnavich}, {Kirshner}, {Crotts}, {Fransson}, {Li}, {Panagia}, {Phillips},
  {Schmidt}, {Sonneborn}, {Suntzeff}, {Wang}, \& {Wheeler}}]{Michael03}
---. 2003, \apj, 593, 809

\bibitem[{{Ng} {et~al.}(2009){Ng}, {Gaensler}, {Murray}, {Slane}, {Park},
  {Staveley-Smith}, {Manchester}, \& {Burrows}}]{Ng09}
{Ng}, C.-Y., {Gaensler}, B.~M., {Murray}, S.~S., {Slane}, P.~O., {Park}, S.,
  {Staveley-Smith}, L., {Manchester}, R.~N., \& {Burrows}, D.~N. 2009, \apjl,
  706, L100

\bibitem[{{Ng} {et~al.}(2008){Ng}, {Gaensler}, {Staveley-Smith}, {Manchester},
  {Kesteven}, {Ball}, \& {Tzioumis}}]{Ng08}
{Ng}, C.-Y., {Gaensler}, B.~M., {Staveley-Smith}, L., {Manchester}, R.~N.,
  {Kesteven}, M.~J., {Ball}, L., \& {Tzioumis}, A.~K. 2008, \apj, 684, 481

\bibitem[{{Panagia}(1999)}]{Panagia99}
{Panagia}, N. 1999, in IAU Symposium, Vol. 190, New Views of the Magellanic
  Clouds, ed. {Y.-H.~Chu, N.~Suntzeff, J.~Hesser, \& D.~Bohlender}, 549

\bibitem[{{Park} {et~al.}(2002){Park}, {Burrows}, {Garmire}, {Nousek},
  {McCray}, {Michael}, \& {Zhekov}}]{Park02}
{Park}, S., {Burrows}, D.~N., {Garmire}, G.~P., {Nousek}, J.~A., {McCray}, R.,
  {Michael}, E., \& {Zhekov}, S. 2002, \apj, 567, 314

\bibitem[{{Park} {et~al.}(2004){Park}, {Zhekov}, {Burrows}, {Garmire}, \&
  {McCray}}]{Park04}
{Park}, S., {Zhekov}, S.~A., {Burrows}, D.~N., {Garmire}, G.~P., \& {McCray},
  R. 2004, \apj, 610, 275

\bibitem[{{Park} {et~al.}(2006){Park}, {Zhekov}, {Burrows}, {Garmire},
  {Racusin}, \& {McCray}}]{Park06}
{Park}, S., {Zhekov}, S.~A., {Burrows}, D.~N., {Garmire}, G.~P., {Racusin},
  J.~L., \& {McCray}, R. 2006, \apj, 646, 1001

\bibitem[{{Park} {et~al.}(2005){Park}, {Zhekov}, {Burrows}, \&
  {McCray}}]{Park05}
{Park}, S., {Zhekov}, S.~A., {Burrows}, D.~N., \& {McCray}, R. 2005, \apjl,
  634, L73

\bibitem[{{Park} {et~al.}(2011){Park}, {Zhekov}, {Burrows}, {Racusin}, {Dewey},
  \& {McCray}}]{Park11}
{Park}, S., {Zhekov}, S.~A., {Burrows}, D.~N., {Racusin}, J.~L., {Dewey}, D.,
  \& {McCray}, R. 2011, \apjl, 733, L35

\bibitem[{{Potter} {et~al.}(2009){Potter}, {Staveley-Smith}, {Ng}, {Ball},
  {Gaensler}, {Kesteven}, {Manchester}, {Tzioumis}, \& {Zanardo}}]{Potter09}
{Potter}, T.~M., {et~al.} 2009, \apj, 705, 261

\bibitem[{{Racusin} {et~al.}(2009){Racusin}, {Park}, {Zhekov}, {Burrows},
  {Garmire}, \& {McCray}}]{Racusin09}
{Racusin}, J.~L., {Park}, S., {Zhekov}, S., {Burrows}, D.~N., {Garmire}, G.~P.,
  \& {McCray}, R. 2009, \apj, 703, 1752

\bibitem[{{Sturm} {et~al.}(2010){Sturm}, {Haberl}, {Aschenbach}, \&
  {Hasinger}}]{Sturm10}
{Sturm}, R., {Haberl}, F., {Aschenbach}, B., \& {Hasinger}, G. 2010, \aap, 515,
  A5

\bibitem[{{Sugerman} {et~al.}(2002){Sugerman}, {Lawrence}, {Crotts}, {Bouchet},
  \& {Heathcote}}]{Sugerman02}
{Sugerman}, B.~E.~K., {Lawrence}, S.~S., {Crotts}, A.~P.~S., {Bouchet}, P., \&
  {Heathcote}, S.~R. 2002, \apj, 572, 209

\bibitem[{{Suzuki} {et~al.}(1993){Suzuki}, {Shigeyama}, \& {Nomoto}}]{Suzuki93}
{Suzuki}, T., {Shigeyama}, T., \& {Nomoto}, K. 1993, \aap, 274, 883

\bibitem[{{van Adelsberg} {et~al.}(2008){van Adelsberg}, {Heng}, {McCray}, \&
  {Raymond}}]{vanAdelsberg08}
{van Adelsberg}, M., {Heng}, K., {McCray}, R., \& {Raymond}, J.~C. 2008, \apj,
  689, 1089

\bibitem[{{Zhekov} {et~al.}(2005){Zhekov}, {McCray}, {Borkowski}, {Burrows}, \&
  {Park}}]{Zhekov05}
{Zhekov}, S.~A., {McCray}, R., {Borkowski}, K.~J., {Burrows}, D.~N., \& {Park},
  S. 2005, \apjl, 628, L127

\bibitem[{{Zhekov} {et~al.}(2006){Zhekov}, {McCray}, {Borkowski}, {Burrows}, \&
  {Park}}]{Zhekov06}
---. 2006, \apj, 645, 293

\bibitem[{{Zhekov} {et~al.}(2009){Zhekov}, {McCray}, {Dewey}, {Canizares},
  {Borkowski}, {Burrows}, \& {Park}}]{Zhekov09}
{Zhekov}, S.~A., {McCray}, R., {Dewey}, D., {Canizares}, C.~R., {Borkowski},
  K.~J., {Burrows}, D.~N., \& {Park}, S. 2009, \apj, 692, 1190

\bibitem[{{Zhekov} {et~al.}(2010){Zhekov}, {Park}, {McCray}, {Racusin}, \&
  {Burrows}}]{Zhekov10}
{Zhekov}, S.~A., {Park}, S., {McCray}, R., {Racusin}, J.~L., \& {Burrows},
  D.~N. 2010, \mnras, 407, 1157

\end{thebibliography}
